\documentclass[12pt,preprint]{aastex}
\usepackage{url}
\usepackage[usenames]{color}
\usepackage{amsmath}
\newcommand{\logg} {\log \textsl{\textrm{g}}}

\newcommand{\Te} {T_{\rm eff}}

\newcommand{\msun} {$M_\odot$}

\newcommand\gta{\lower 0.5ex\hbox{$\buildrel > \over \sim\ $}} 
\newcommand\lta{\lower 0.5ex\hbox{$\buildrel < \over \sim\ $}} 
\newcommand{\nh} {{\rm H}/{\rm He}}
\newcommand{\halpha} {$\rm{H}{\alpha}$}
\newcommand{\hbeta} {$\rm{H}{\beta}$}

\newcommand{\heii} {He {\sc ii}}

\newcommand{\civ} {C {\sc iv}}

\shortauthors{Bergeron et al.}
\shorttitle{MCT White Dwarfs in the Southern Hemisphere}

\begin{document}

\title {Hot Degenerates in the MCT Survey. III. A Sample of White\\
        Dwarf Stars in the Southern Hemisphere}

\author{P. Bergeron\altaffilmark{1}, F. Wesemael\altaffilmark{1}, G. Fontaine\altaffilmark{1}, 
      R. Lamontagne\altaffilmark{1}, S. Demers\altaffilmark{1}, A. B\'edard} 
\affil{D\'epartement de Physique, Universit\'e de Montr\'eal,
      Montr\'eal, Qu\'ebec H3C 3J7, Canada\\ 
      bergeron, lamont, demers, bedard@astro.umontreal.ca}
\author{M.-J. Gingras} 
\affil{Department of Physics and Astronomy, University of Waterloo, Waterloo, ON N2L 3G1, Canada\\
       m3gingras@uwaterloo.ca}
\author{S. Blouin} 
\affil{Los Alamos National Laboratory, Los Alamos, NM 87545, USA\\
       sblouin@lanl.gov}
\author{M.J. Irwin\altaffilmark{1}}
\affil{Royal Greenwich Observatory, Madingley Road, Cambridge, CB3 0EZ,\\
       United Kingdom\\
       mike@ast.cam.ac.uk}
\author{S.O. Kepler\altaffilmark{1}}
\affil{Instituto de Fisica, Universidade Federal do Rio Grande do Sul, RS, Brazil\\
       kepler@if.ufrgs.br}

\altaffiltext{1}{Visiting Astronomer, Cerro Tololo Interamerican
Observatory, National Optical Astronomical Observatories, which is
operated by AURA, Inc., under contract with the National Science
Foundation.}

\begin{abstract}

We present optical spectra of 144 white dwarfs detected in the
Montreal-Cambridge-Tololo (MCT) colorimetric survey, including 120 DA,
12 DB, 4 DO, 1 DQ, and 7 DC stars. We also perform a model atmosphere
analysis of all objects in our sample using the so-called
spectroscopic technique, or the photometric technique in the case of
DC white dwarfs. The main objective of this paper is to contribute to
the ongoing effort of confirming spectroscopically all white dwarf
candidates in the {\it Gaia} survey, in particular in the southern
hemisphere. All our spectra are made available in the Montreal White
Dwarf Database.

\end{abstract}

\section{Introduction}

Large-scale colorimetric surveys are an important source of new hot
subluminous objects that can form the basis of many different types of
investigations. For instance, the Palomar-Green (PG) survey
\citep{PG86} has yielded a complete sample of white dwarfs from which
the luminosity function of hot white dwarfs could be derived (see
\citealt{Liebert2005} for DA stars, and \citealt{Bergeron2011} for DB
stars). In addition, the PG survey increased considerably the number
of objects in some sparsely populated classes (e.g., the DBA stars),
as well as revealed the existence of entirely new kinds of objects
(e.g., the very hot PG 1159 stars).  The PG yield of subluminous
objects reawakened interest in colorimetric surveys, which have a long
and distinguished history
\citep{Humason1947,Feige1958,Haro1962,Lanning1973} in white dwarf
astronomy. The following years have witnessed a flurry of activity in
this area, with at least four analogs of the PG survey proceeding
concurrently: the Kiso survey \citep{Noguchi1980, Kondo1984}; the
Montreal-Cambridge-Tololo (MCT) survey \citep{Demers1986}; the
Edinburgh-Cape survey \citep{Stobie1987,Stobie1992,Stobie1997}; and
the Homogeneous Bright Quasars Survey, a Key Project carried out at
ESO \citep[][see also
  \citealt{Cristiani1995}]{Gemmo1993,Gemmo1995}. At the same time, the
yield of new degenerates identified in objective-prism surveys like
the Hamburg/ESO and Hamburg Quasar Surveys
\citep{Reimers1996,Reimers1998,Homeier1998} and the earlier Case
Low-Dispersion Northern Sky Survey \citep{Wagner1988} cannot be
understated.

Of course, these old colorimetric surveys have now been completely
superseded by the large Sloan Digital Sky Survey (SDSS), at least in
the northern hemisphere, which identified well over 30,000 white
dwarfs in the Data Release 14 \citep{Kepler2019}. In parallel, the
{\it Gaia} Data Release 2 \citep[][hereafter GaiaHRD]{Gaia2018} has
provided precise astrometric and photometric data for $\sim$260,000
high-confidence white dwarf candidates
\citep{GentileFusillo2019}. Although the spectroscopic follow-up of
the {\it Gaia} sample is still in progress (see, e.g.,
\citealt{Tremblay2020}), many white dwarf candidates still require
spectroscopic observations and confirmations, particularly in the
southern hemisphere.

As part of this global effort, we decided to publish the spectroscopic
results from the MCT survey, long overdue, and to make all of our
spectra available in the Montreal White Dwarf
Database\footnote{http://montrealwhitedwarfdatabase.org/}
\citep[MWDD;][]{Dufour2017}. We thus present in Section 2 a sample of 144 white
dwarfs observed spectroscopically, several of which have previously
been identified in other surveys, and discussed in the
literature. These spectra are then analyzed in Section 3, and the
global properties of our sample are discussed in Section 4.

\section{Observations}

\subsection{The Photographic Survey}

A detailed description of the photographic part of the MCT survey has
been presented by \citet{Demers1986}. In essence, doubly-exposed IIa O
plates exposed through $U$ and $B$ filters were obtained at the CTIO
Curtis Schmidt telescope. Exposure times are adjusted so as to yield
comparable image sizes for objects with $(U-B)\sim -0.5$. The
photographic plates were then measured using the Automatic Plate
Measuring System (APM) at Cambridge \citep{Kibblewhite1984}. An
internal magnitude scale was defined for both images, and the internal
magnitudes were then transformed into $B$ and $U$ magnitudes either
through published photoelectric sequences, or through calibrated CCD
frames. As a general rule, we concentrated on the bluest objects,
those with $(U-B)\leq -0.6$, although some redder objects, with
$-0.6\leq (U-B)\leq -0.4$ have been occasionally observed. All blue
candidates were first visually inspected to weed out interlopers
(e.g., $U$ image contaminated by the $B$ of a nearby star, stars near
the edge of the plate, etc.).

\begin{figure}[p]
\centering
\includegraphics[width=0.9\linewidth]{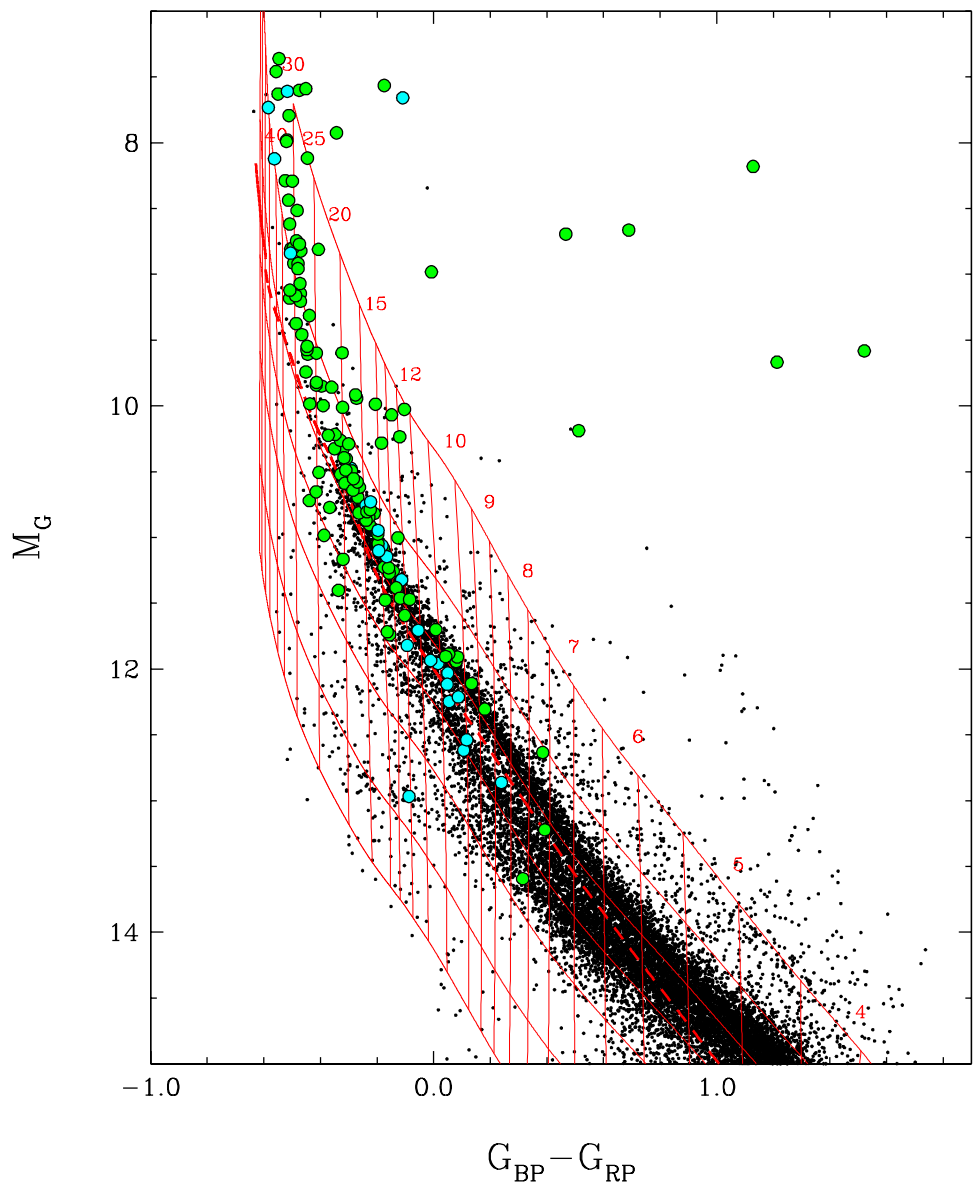}
\caption{Gaia color-magnitude diagram for the 100 pc sample (small
  dots) drawn from the MWDD; 
  the white dwarfs in the MCT survey with H and He atmospheres are
  indicated by filled green and cyan dots, respectively. Red solid
  lines show the cooling sequences for CO-core and pure H atmosphere
  white dwarf models with 0.2, 0.4, 0.6, 0.8, 1.0, 1.2, and 1.3
  \msun\ (from top to bottom); the dashed line shows the sequence for
  0.6 \msun\ pure He atmosphere models; $\Te$ values are indicated in units of $10^3$ K.\label{color_mag}}
\end{figure}

Here we go one step further and take advantage of the {\it Gaia}
photometric and astrometric data (EDR3; \citealt{Brown2021,
  Lindegren2021}) to confirm the nature of the objects identified in
the MCT survey. In Figure \ref{color_mag} we present the $M_G$ vs
$(G_{\rm BP}-G_{\rm RP})$ color-magnitude diagram for the white dwarfs
within 100 pc from the Sun identified in the MWDD, together with the
white dwarfs identified in our survey. Different color symbols are
used to distinguish H- and He-dominated atmospheres. Hot subdwarfs and
other contaminants have been removed from the sample and can be made
available upon request. Also shown are theoretical color sequences
taken from our publicly available Web site\footnote{See
  http://www.astro.umontreal.ca/$\sim$bergeron/CoolingModels.}. The
observed sequence follows nicely the color track near $\sim$0.6 \msun,
with many hot white dwarfs reaching the Rayleigh-Jeans limit near
$(G_{\rm BP}-G_{\rm RP})\sim0.5$, and several red objects whose colors
are obviously contaminated by the presence of an M-dwarf companion are
also present in the sample (see below).  As expected, most white
dwarfs in our sample are found at high effective temperatures
($\Te\gtrsim10,000$ K), although several cooler objects have also been
uncovered.

\subsection{Spectroscopic Observations}

Follow-up observations of the blue candidates in the MCT survey have
been secured, beginning as long ago as 1985. Most of the observations
were obtained at the CTIO 1.5-m and 4-m telescopes, with a variety of
instrumental setups.  Early runs at the 1.5-m made use of the UV SIT,
RCA CCD no.~5, and GE CCD EPI no.~4, all in conjunction with the
Cassegrain spectrograph.  From 1988 to 1994, however, data has been
gathered with the GEC CCD no.~10. For our last run (Nov. - Dec. 1995),
the detector was upgraded to a thinned Loral $1200\times 400$ device.
The spectral resolution achieved is 6.5--13 \AA. At the CTIO 4-m
telescope, data were gathered with the RC spectrograph, Air Schmidt or
Folded Schmidt camera, and either the GEC no.~11 or the TI no.~1 CCD
with a resolution of 6--7 \AA. In addition, two runs were secured in
1987 and 1989 at the Las Campanas 2.5-m telescope, equipped with a 2D
Frutti detector, with a spectral resolution of 3 \AA. Finally, a
handful of spectra extending into the red were secured through service
observing at the La Palma Observatory with the double
spectrograph. Further details of our spectroscopic follow-up can be
found in \citet{Lamontagne2000}.

We provide in Table 1 the list of objects that will be analyzed in the
next section, including the MCT name, the 1950 position, the
photographic values of $B_{\rm ph}$ and $(U-B)_{\rm ph}$, as well as
the date, telescope, and spectral resolution (FWHM) for each spectrum.

The optical spectra for the 120 DA white dwarfs identified in
our survey are displayed in Figure \ref{spec_DA} in order of
decreasing effective temperature. At the level of resolution of our
spectroscopic observations, we could not identify any DAZ stars in our
sample. Already obvious in this figure, however, is the contamination
from an M-dwarf companion in several objects (see, e.g., MCT
0309$-$2730, MCT 2108$-$4310, MCT 2259$-$321). 

\begin{figure}[p]
\centering
\includegraphics[width=0.9\linewidth]{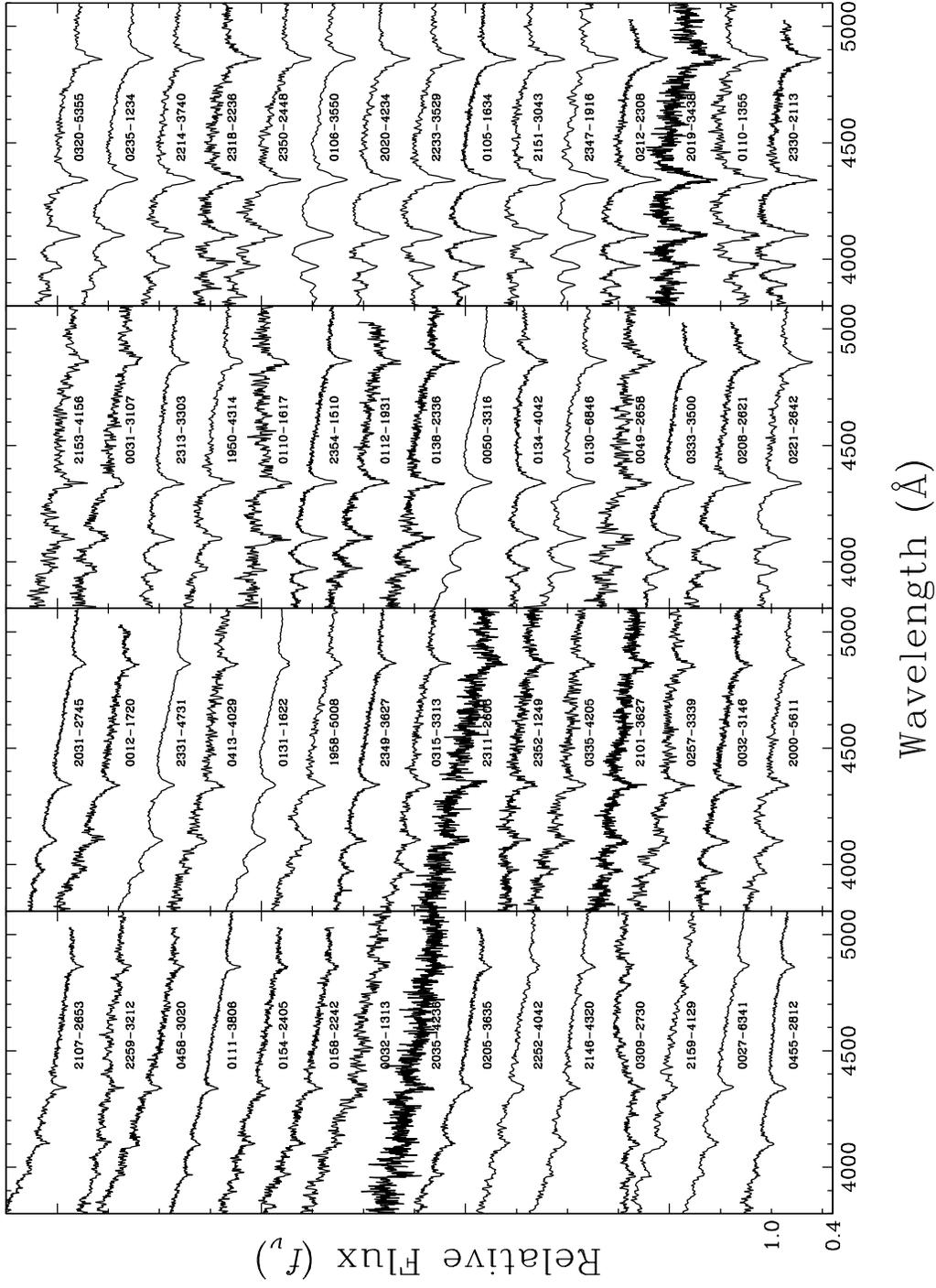}
\caption{Optical (blue) spectra for our complete sample of DA stars. The
spectra are normalized at 4500 \AA\ and shifted vertically from each
other by a factor of 0.5 for clarity. The effective temperature
decreases from upper left to bottom right.\label{spec_DA}}
\end{figure}

\begin{figure}[p]
\figurenum{2}
\centering
\includegraphics[width=0.9\linewidth]{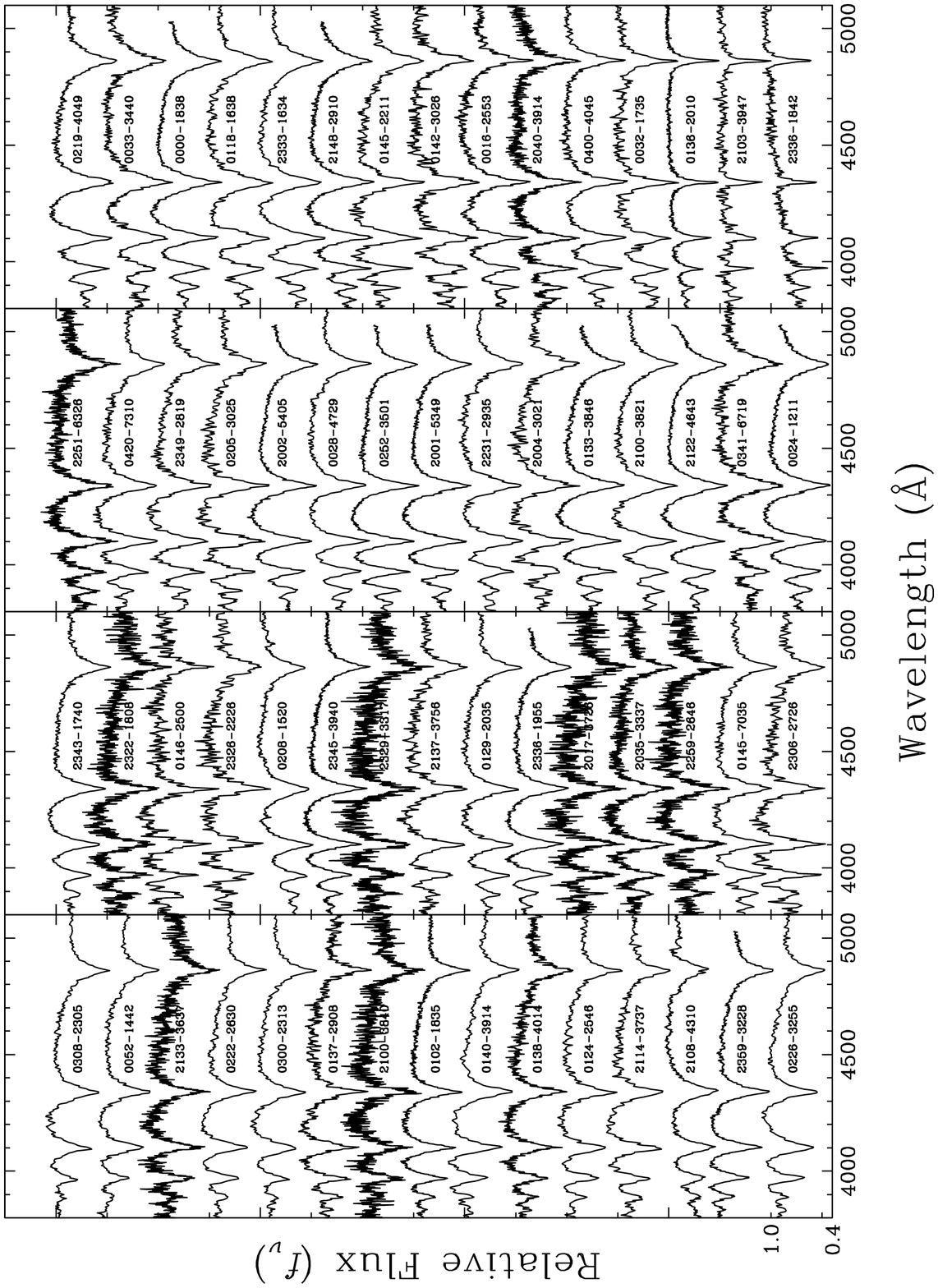}
\caption{(Continued)}
\end{figure}

\begin{figure}[p]
\centering
\includegraphics[width=0.9\linewidth]{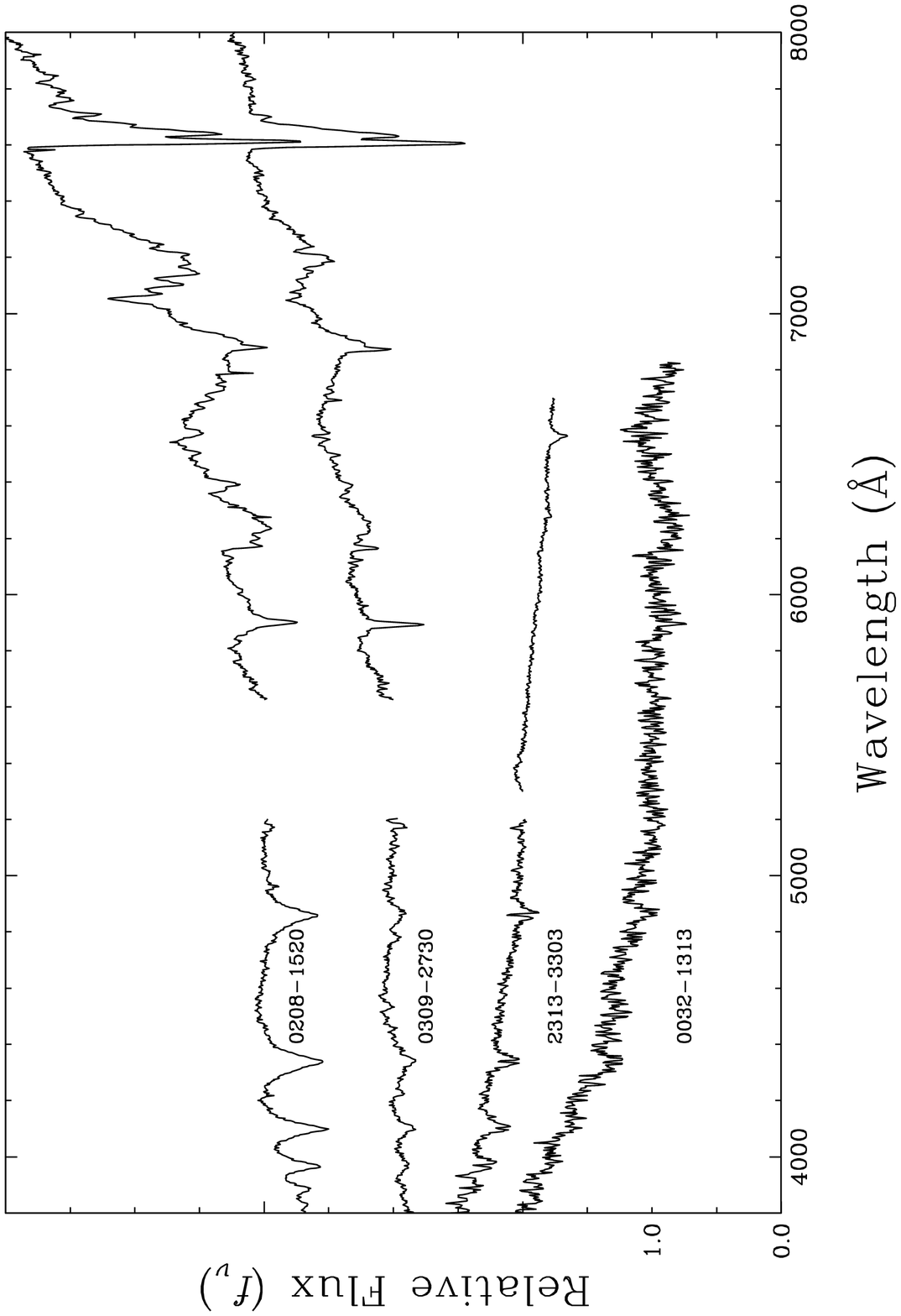}
\caption{Composite spectra of DA+M-dwarf systems from Figure \ref{spec_DA}
  for which red optical spectra are also available.  The secondary
  star for 0208$-$1520 appears be a M3 or M4 V companion, while for
  0309$-$2730, the secondary star appears be a M0 or M1 V
  companion.\label{spec_composite}}
\end{figure}

Composite DA+M-dwarf systems that also have red spectra are
displayed in Figure Figure \ref{spec_composite}. From top to bottom:
The blue spectrum of MCT 0208$-$1520 suggests the presence of
additional features near MgH $\lambda$4780. This is confirmed by the
spectrum longward of 5500 \AA, which rises steeply in the red, and
which appears to match that of an M3 or M4 V companion. The blue
spectrum of MCT 0309$-$2730 appears disturbed, and the presence of an
unresolved companion is confirmed by the red spectrum, which matches
that of an early-type M V star, perhaps M0 or M1. The blue spectrum of
MCT 2313$-$3303 displayed in Figure \ref{spec_composite} is clearly
reminiscent of that of RE 2013+400 \citep{Barstow1993} and RE
0720$-$318 \citep{Barstow1995}, where the cores of the Balmer lines
are filled with reprocessed emission by the irradiated hemisphere of
an M dwarf companion. In MCT 2313$-$3303, emission cores are seen in
H$\beta$ and upward, while our red spectrum shows no emission at
H$\alpha$. However, the latter was secured two days later than the
blue spectrum, and the emission might be variable (as it is in RE
2013+400). The red coverage of the bottom object in Figure
\ref{spec_composite}, MCT 0032$-$1313, is not as good as that of the
two objects at the top of the figure, but it is sufficient to detect
the presence of the M-dwarf companion.

\begin{figure}[p]
\centering
\includegraphics[width=0.9\linewidth]{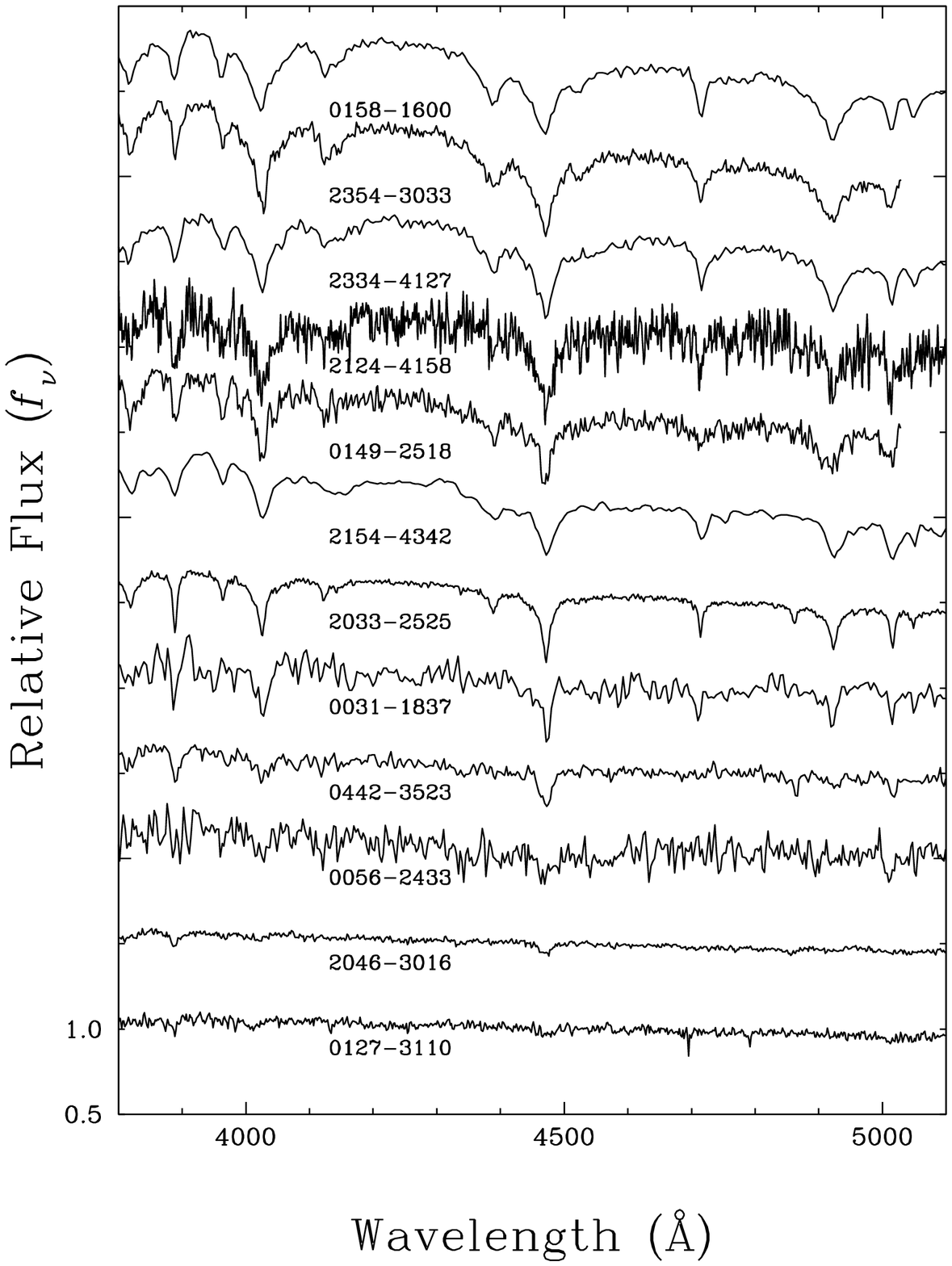}
\caption{Optical (blue) spectra for our complete sample of DB and DBA
  stars. The spectra are normalized at 4500 \AA\ and shifted
  vertically from each other by a factor of 0.5 for clarity. The
  effective temperature decreases from top to bottom.\label{spec_DB}}
\end{figure}

The optical spectra for the 12 DB white dwarfs identified in
our survey are displayed in Figure \ref{spec_DB} in order of
decreasing effective temperature. In three of these objects (MCT
0442$-$3523, MCT 2033$-$2525, and MCT 2046$-$3016), \hbeta\ can be
detected, which makes them DBA stars. Given our spectral coverage,
we cannot exclude that some of the DB white dwarfs displayed here could
also show an \halpha\ absorption feature. 

\begin{figure}[p]
\centering
\includegraphics[width=0.9\linewidth]{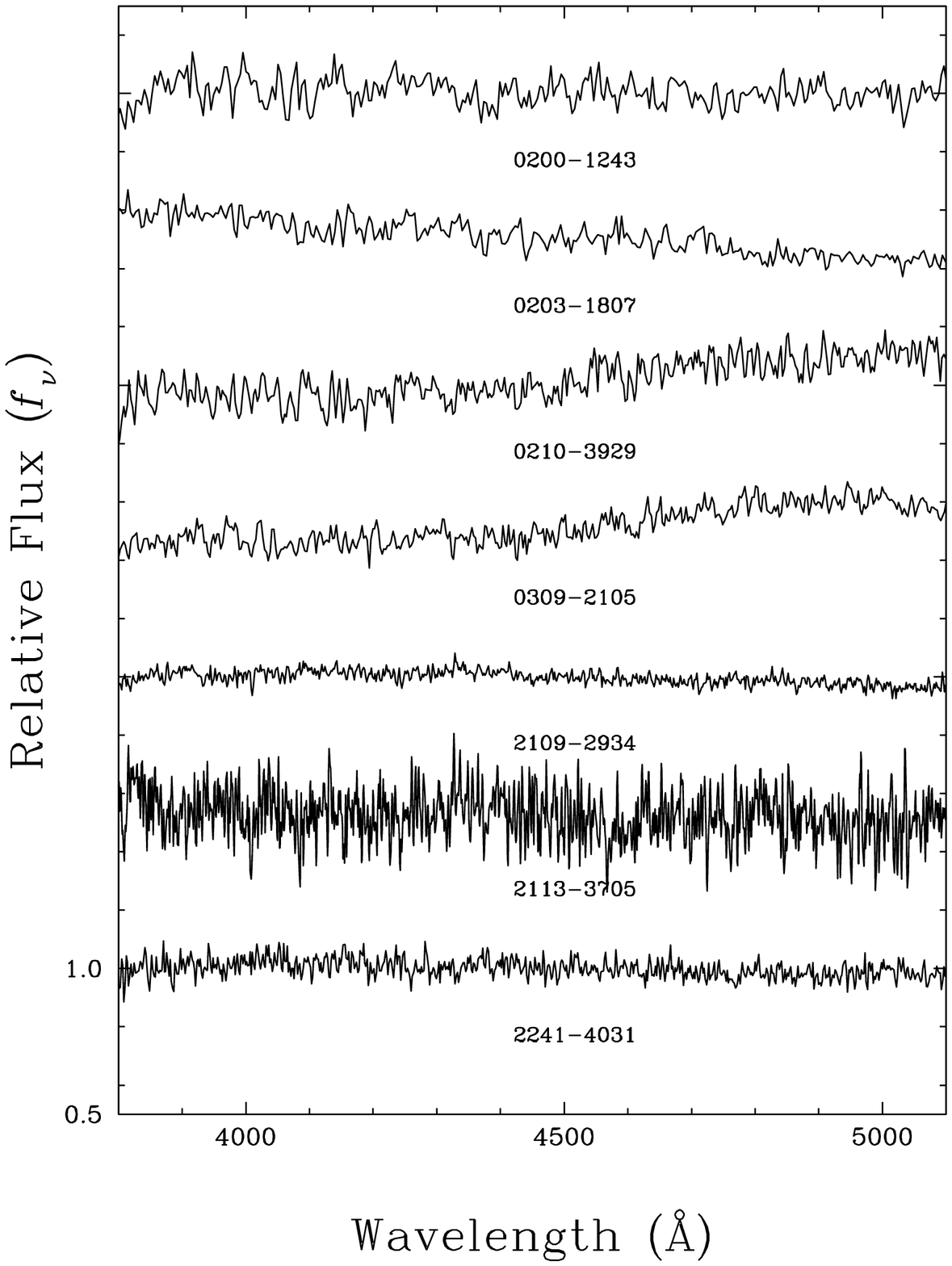}
\caption{Optical (blue) spectra for our complete sample of DC stars,
  ordered by right ascension. The spectra are normalized at 4500
  \AA\ and shifted vertically from each other by a factor of 0.5 for
  clarity.\label{spec_DC}}
\end{figure}

The optical spectra for the 7 DC white dwarfs identified in our
survey are displayed in Figure \ref{spec_DC} in order of right
ascension. Our photometric fits discussed below indicate that all
these stars are hot enough to show hydrogen lines if they were DA
stars, hence they must have He-dominated atmospheres. However, due to
the lack of spectral coverage in the red, we cannot exclude that some
of these objects could display a broad and shallow \halpha\ absorption
feature, characteristic of the so-called He-rich DA white dwarfs
\citep{Rolland2018}. 

\begin{figure}[p]
\centering
\includegraphics[width=0.9\linewidth]{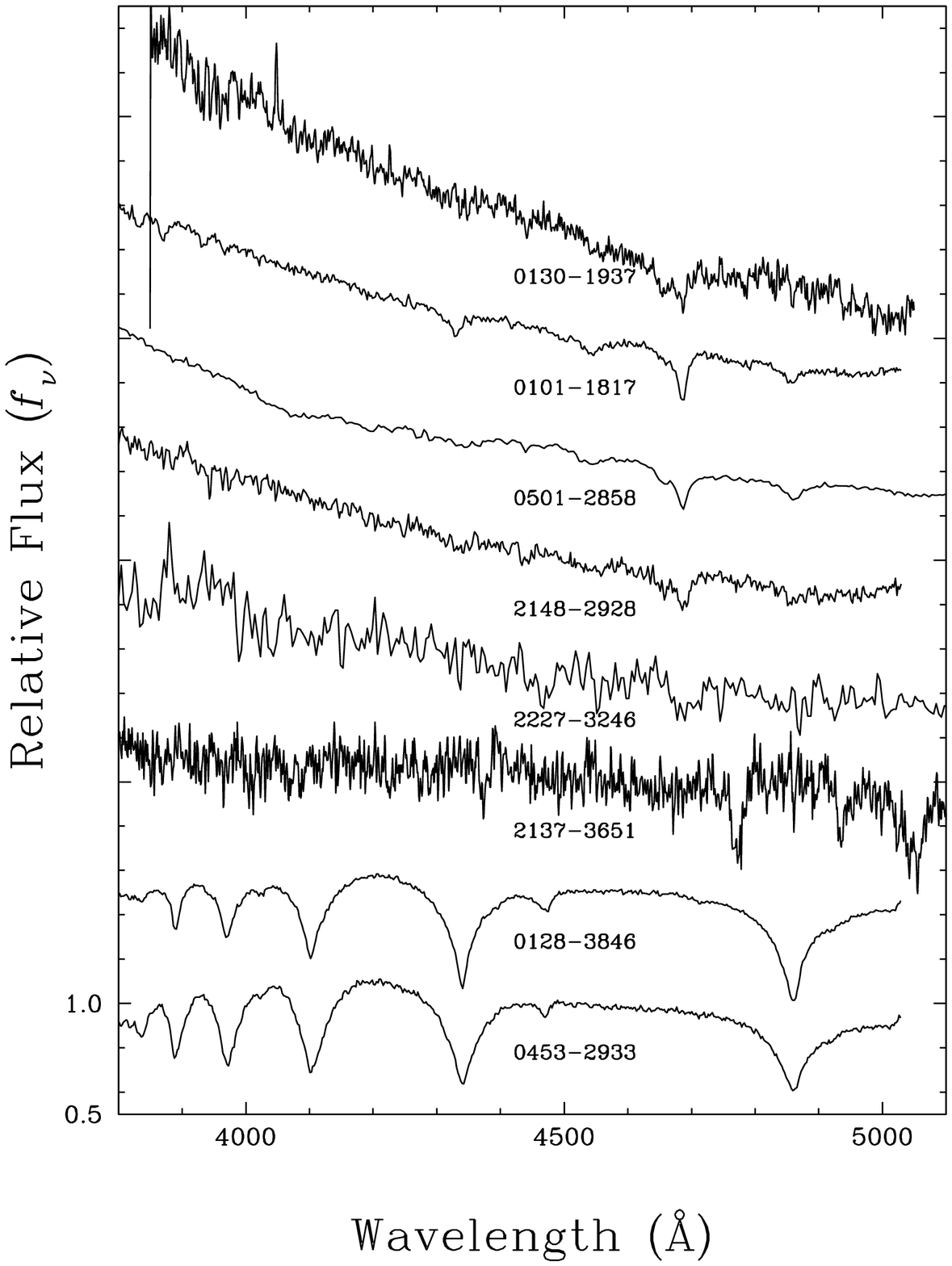}
\caption{Optical (blue) spectra for miscellaneous white dwarfs in our
  sample; all spectra are normalized at 4500 \AA\ and shifted
  vertically from each other by a factor of 0.5 for clarity. From top
  to bottom: MCT 0130$-$1937, a PG 1159 star already discussed in
  \citet{Demers1990}; MCT 0101$-$1817, MCT 0501$-$2858, MCT
  2148$-$2928, and MCT 2227$-$3246, four DO stars analyzed in this
  paper; MCT 2137$-$3651, the only DQ star in our sample, also
  analyzed in this paper; MCT 0128$-$3846 and MCT 0453$-$2933, two
  peculiar DAB stars analyzed in detail by
  \citet{Wesemael1994}.\label{spec_other}}
\end{figure}

Finally, we show in Figure \ref{spec_other} optical spectra for
white dwarfs in our sample with other spectral types. The top object,
MCT 0130$-$1937, is a PG 1159 star already discussed at length in
\citet{Demers1990}, while the bottom two white dwarfs, MCT 0128$-$3846
and MCT 0453$-$2933, correspond to peculiar DAB stars analyzed in
detail by \citet{Wesemael1994}. The analysis of these 3 white dwarfs
will not be repeated here. MCT 0101$-$1817, MCT 0501$-$2858, MCT
2148$-$2928, and MCT 2227$-$3246, are DO stars, while MCT 2137$-$3651
is the only DQ white dwarf identified in our sample. The spectroscopic analysis
of these objects is described below.

\begin{figure}[p]
\centering
\includegraphics[width=0.9\linewidth]{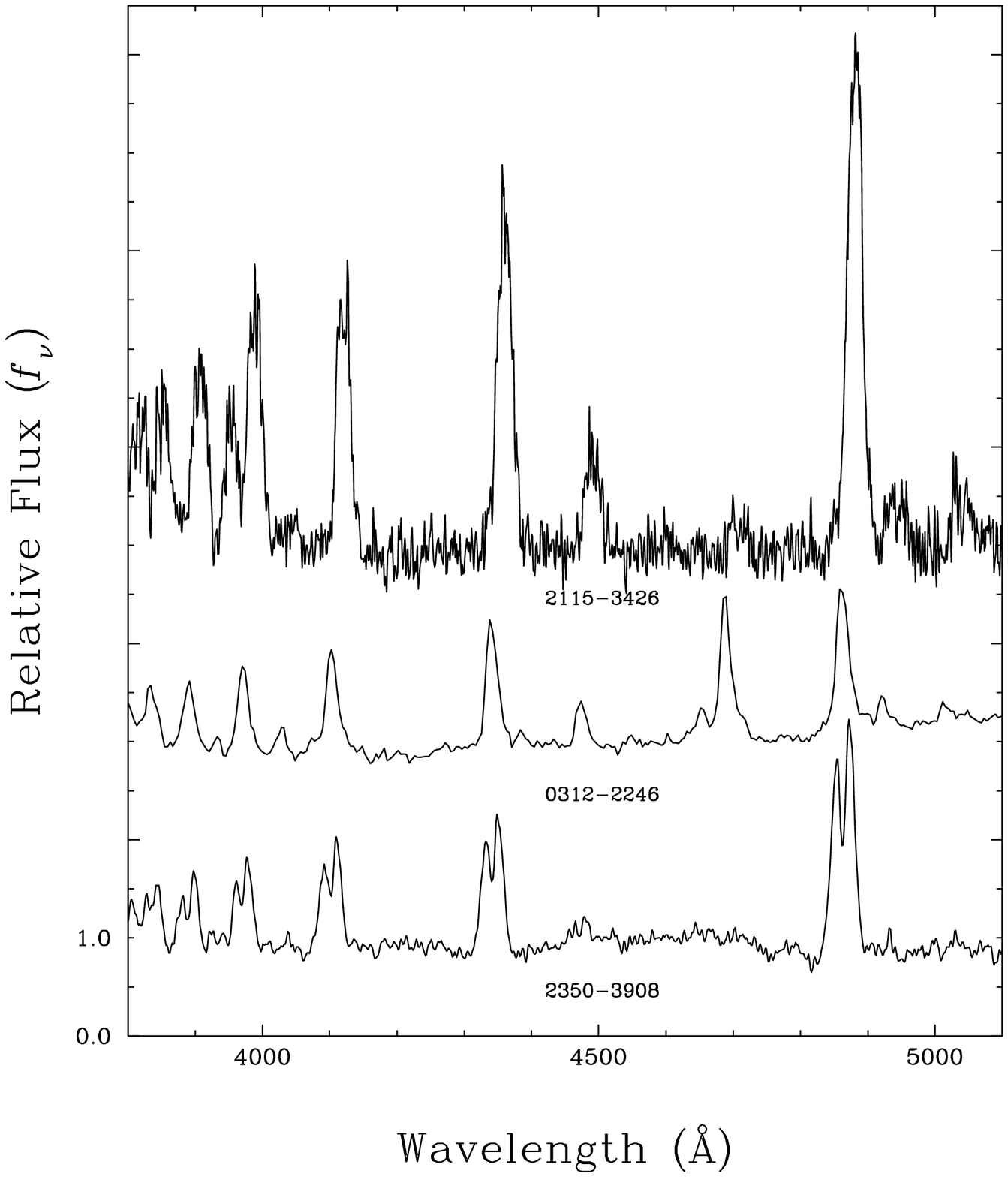}
\caption{Optical (blue) spectra for the cataclysmic variables (MCT
  0312$-$2246 and MCT 2115$-$3426) and the nova (MCT 2350$-$3908)
  identified in our survey.\label{spec_CV}}
\end{figure}

Although not analyzed here, we show for completeness in Figure
\ref{spec_CV} the spectra of objects with strong emission
lines. According to Simbad, MCT 0312$-$2246 and MCT 2115$-$3426 are
cataclysmic variables, while MCT 2350$-$3908 is a nova.

\section{Analysis of the Sample}

In this section, we present a spectroscopic (or photometric) analysis
of all the white dwarfs identified in the MCT survey (with the
exception of three objects already analyzed elsewhere as mentioned above). Given the scope of our
paper, we restrict ourselves to our own spectroscopic data, and thus,
we do not attempt to obtain the most reliable atmospheric parameters
for each object, nor will we cross-match our solutions with those
already published in the literature.

The atmospheric parameters ($\Te$ and $\logg$) for the DA stars in our
sample can be measured using the spectroscopic technique described in
\citet[][]{Bergeron1992}, with various improvements discussed in
\citet{Gianninas2011}. Briefly, the normalized Balmer line profiles
are compared with the predictions of detailed model atmospheres,
properly convolved with the instrumentation profile, and the best
fitting $\Te$ and $\logg$ values are obtained using a $\chi^2$
minimization technique. The model atmospheres used here are similar to
those presented in \citet{Tremblay2011} and references therein, with
the exception that the non-LTE models at high effective temperatures
($\Te>40,000$~K) have been replaced with those calculated by
\citet{Bedard2020}. Convective energy transport has been included in
our calculations using the mixing-length theory with the
ML2/$\alpha=0.7$ convective efficiency. It is a well known problem
that for DA stars close to the region where the Balmer lines reach
their maximum strength ($\Te\sim13,000$~K), there exist two solutions,
one of each side of this maximum \citep{BergeronZZ1995}. To
distinguish between these cool and hot spectroscopic solutions, we
rely on the photometric approach, described further below, to estimate
$\Te$ using the {\it Gaia} astrometric and photometric data.

Our best fits for the DA stars in the MCT sample are displayed in Figure
\ref{fit_DA1} in order of right ascension. The theoretical profiles
shown in green are contaminated by the presence of an M-dwarf
companion, and these lines are not included in the fit. Note that
the 3D hydrodynamical $\Te$ and $\logg$ corrections of
\citet{Tremblay2013} have been applied to the solutions
displayed in these figures (see Section 4).

\begin{figure}[p]
\centering \includegraphics[width=0.9\linewidth]{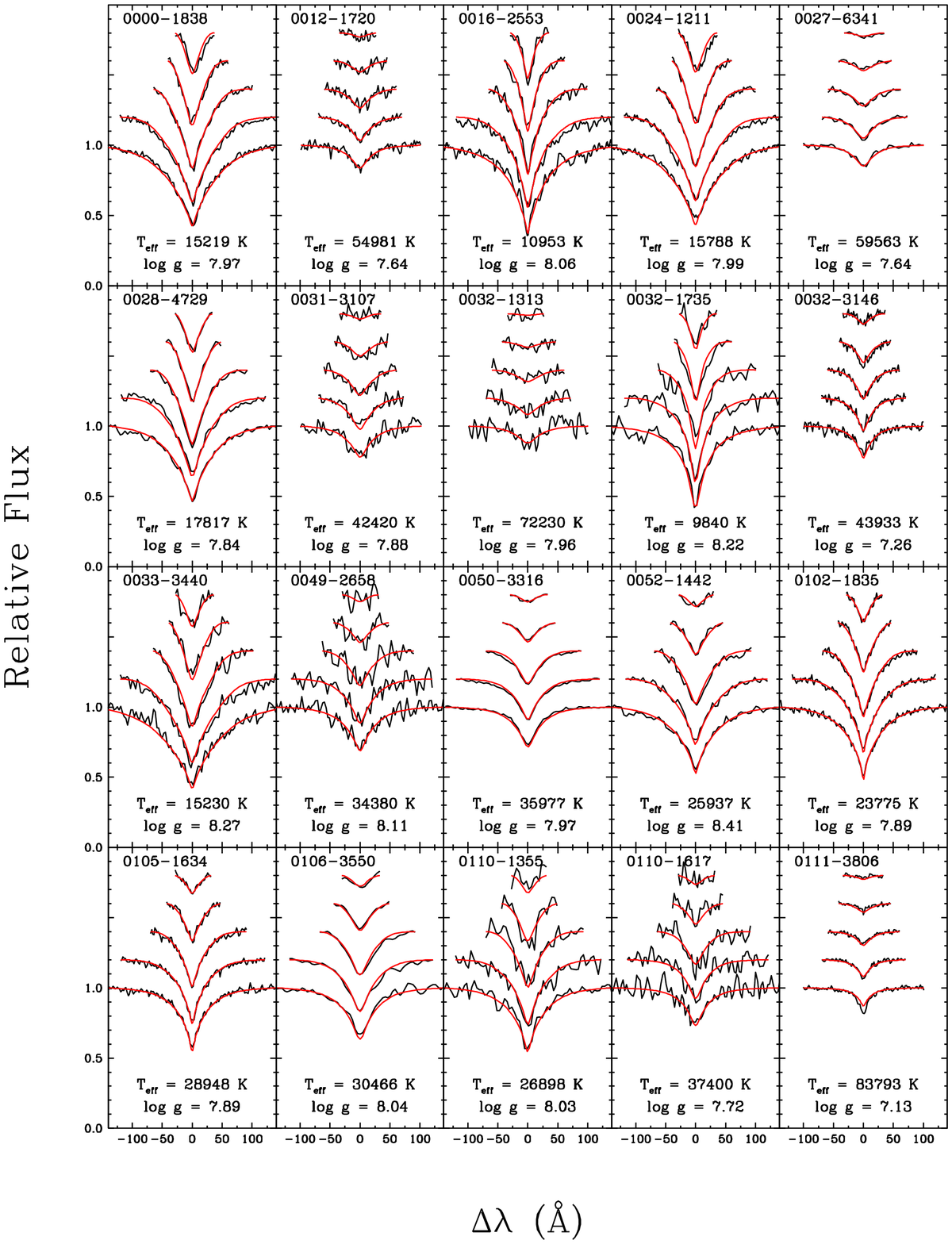}
\caption{Model fits (red) to the individual Balmer line profiles
  (black) of the DA white dwarfs in our sample in order of right
  ascension. The lines range from \hbeta\ (bottom) to H8 (top), each
  offset by a factor of 0.2 (those shown in green are not included in
  the fit). The best-fit $\Te$ and $\logg$ values are indicated at the
  bottom of each panel; note that the 3D hydrodynamical $\Te$ and
  $\logg$ corrections have been applied to the solutions displayed
  here.)\label{fit_DA1}}
\end{figure}

\begin{figure}[p]
\figurenum{8}
\centering
\includegraphics[width=0.9\linewidth]{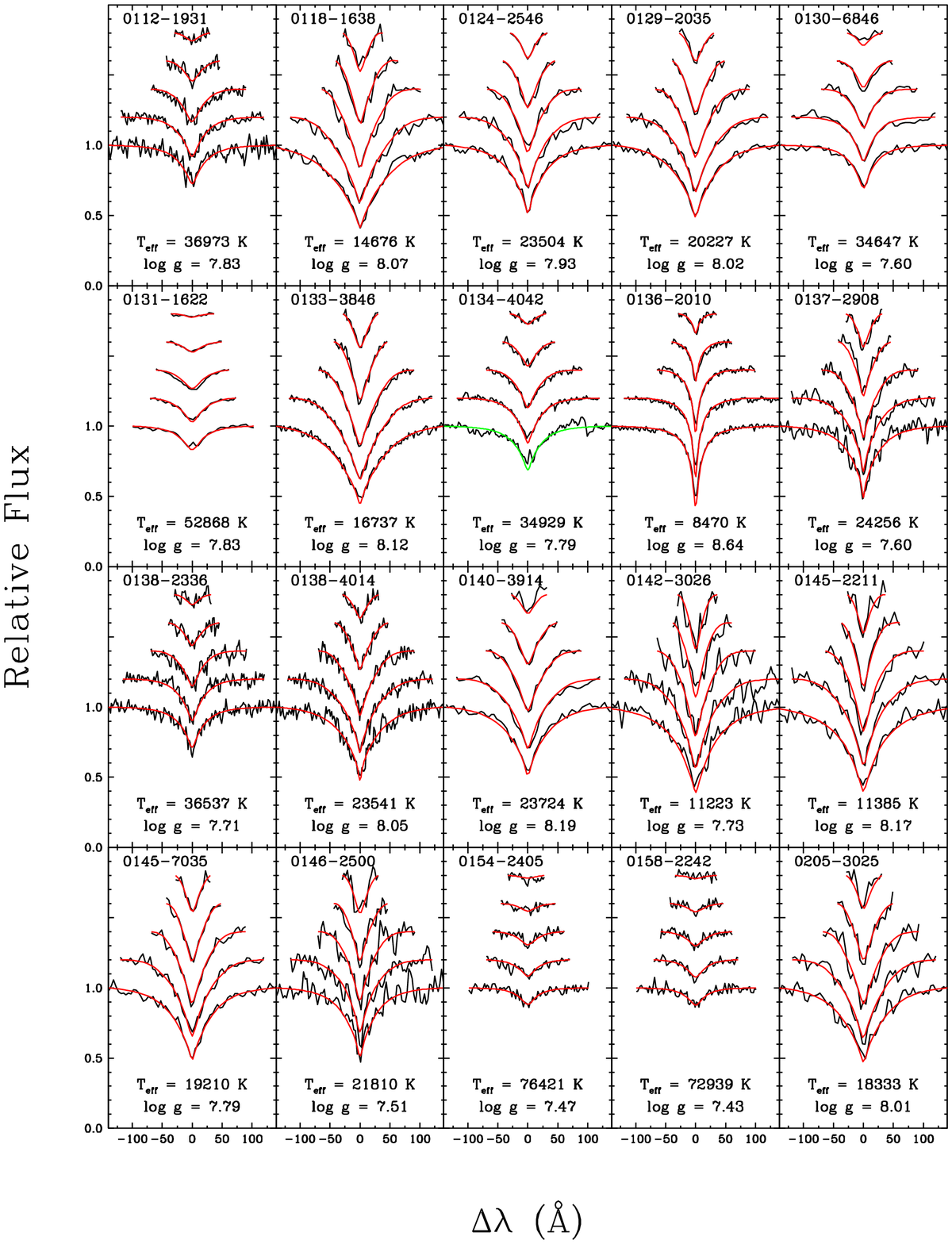}
\caption{(Continued)}
\end{figure}

\begin{figure}[p]
\figurenum{8}
\centering
\includegraphics[width=0.9\linewidth]{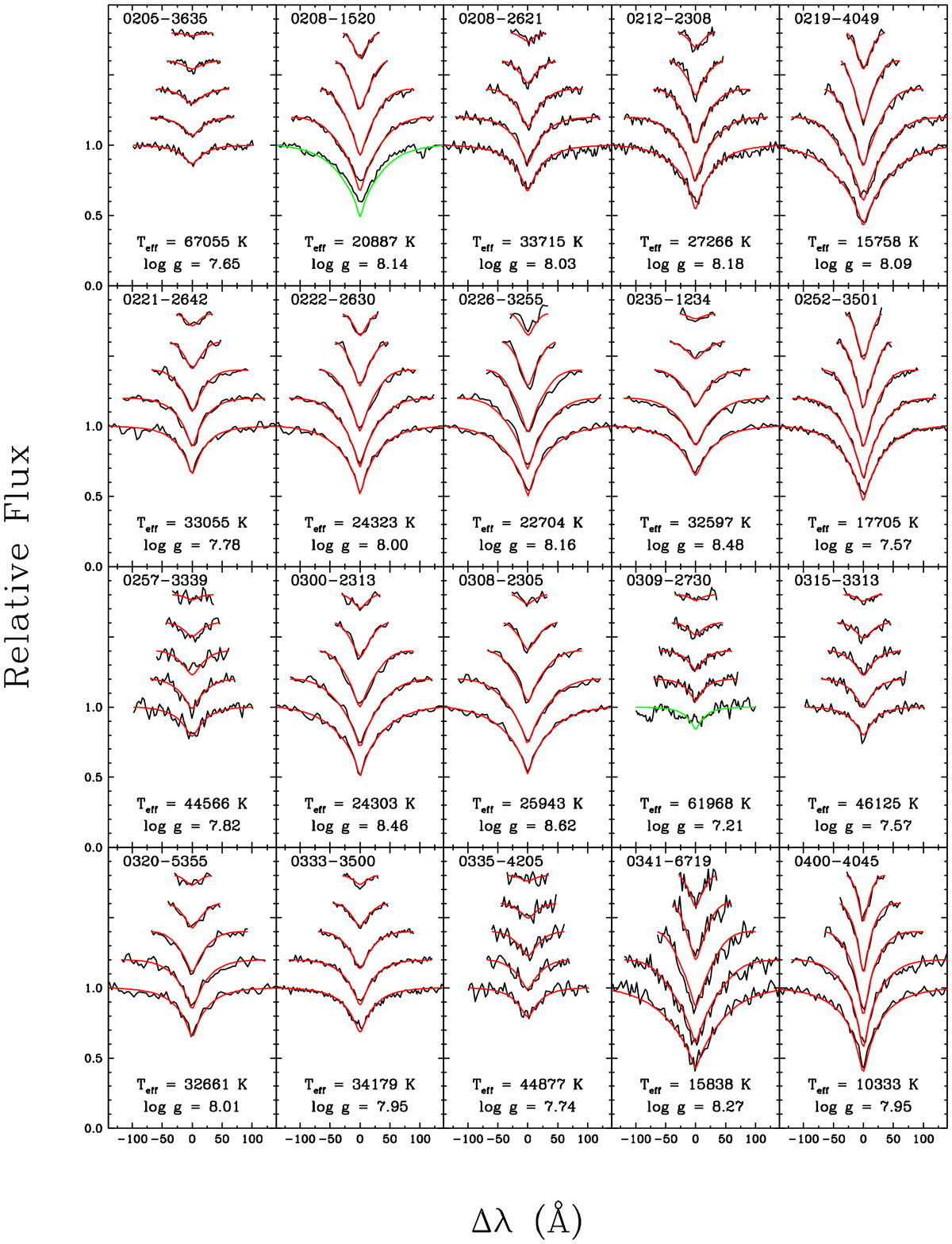}
\caption{(Continued)}
\end{figure}

\begin{figure}[p]
\figurenum{8}
\centering
\includegraphics[width=0.9\linewidth]{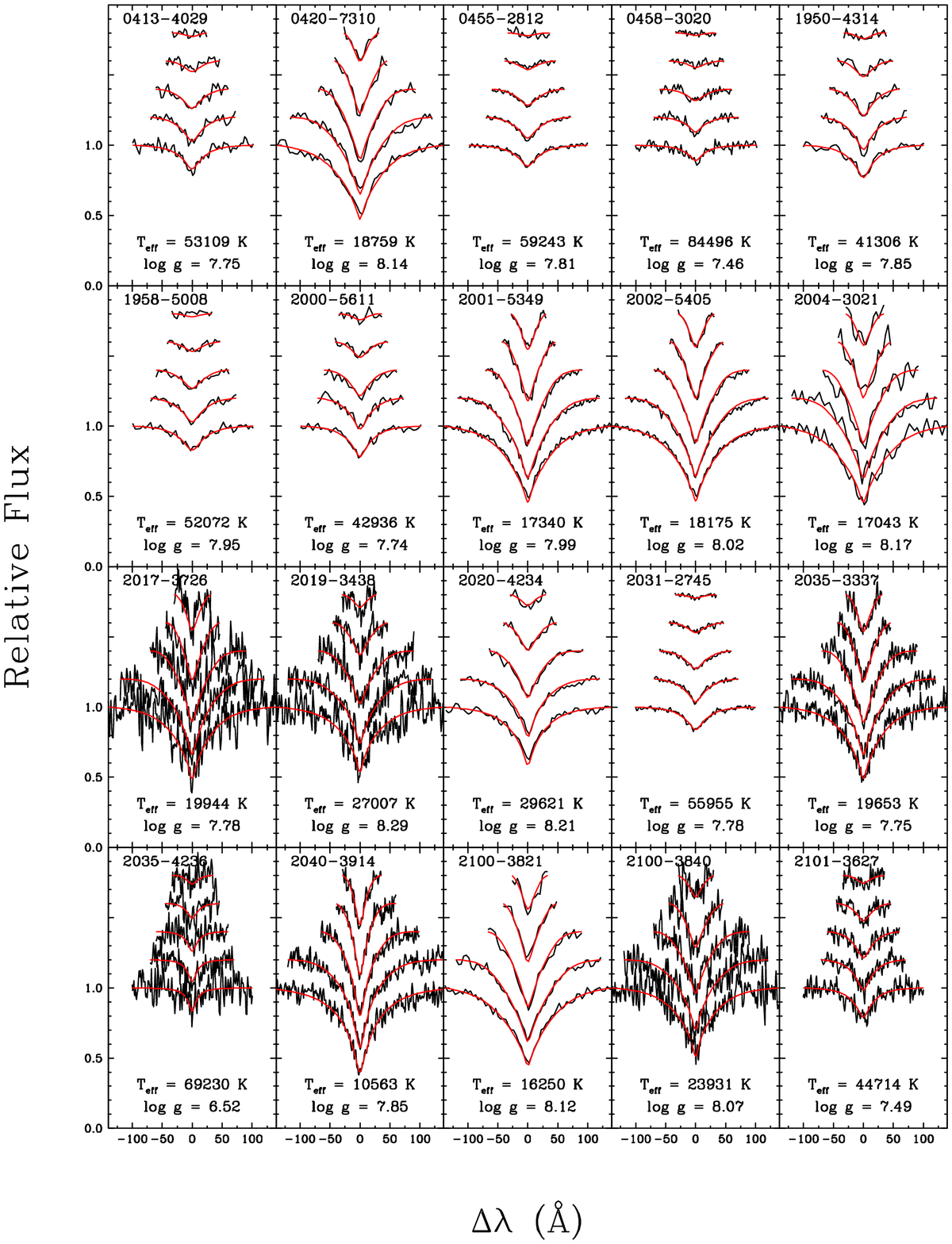}
\caption{(Continued)}
\end{figure}

\begin{figure}[p]
\figurenum{8}
\centering
\includegraphics[width=0.9\linewidth]{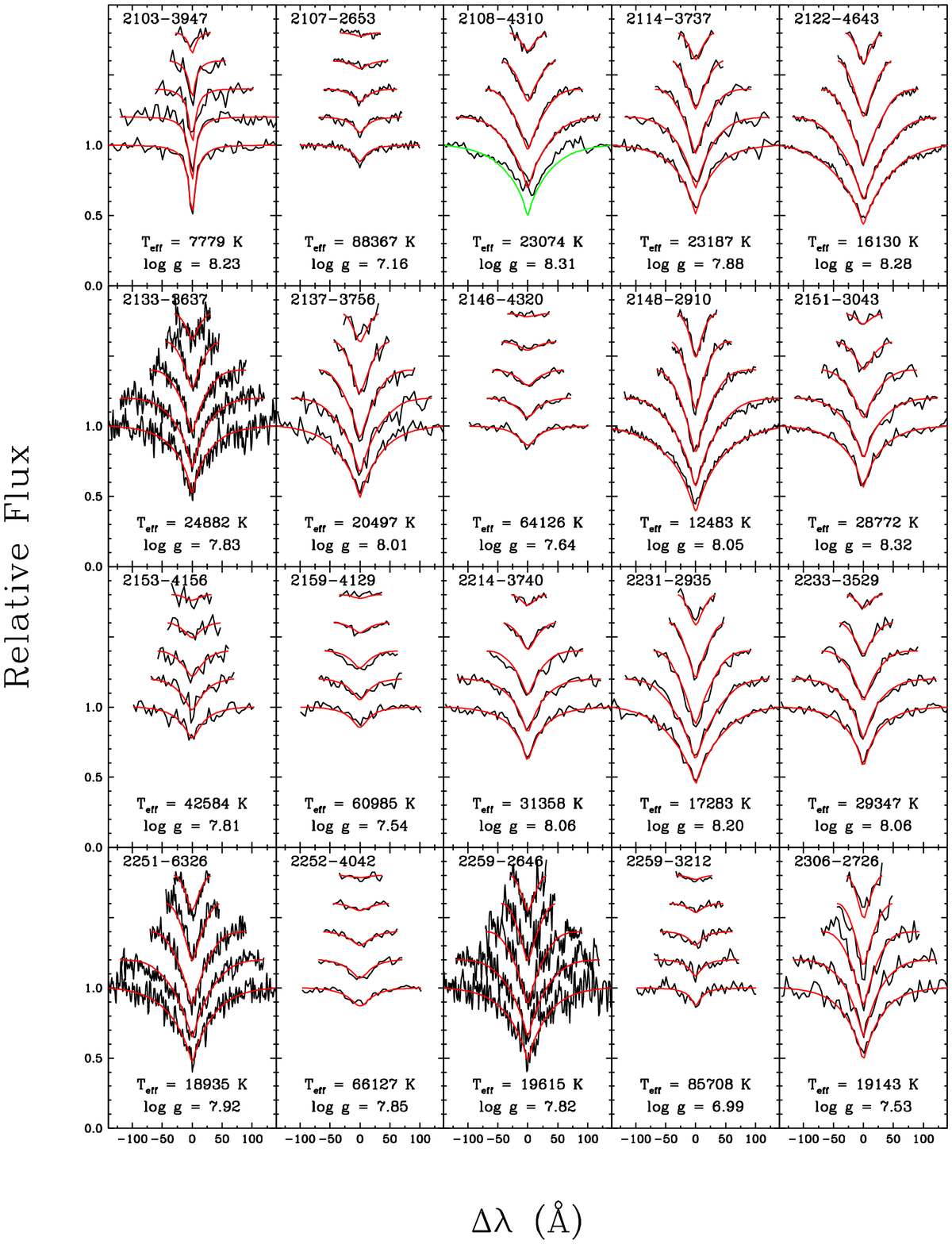}
\caption{(Continued)}
\end{figure}

\begin{figure}[p]
\figurenum{8}
\centering
\includegraphics[width=0.9\linewidth]{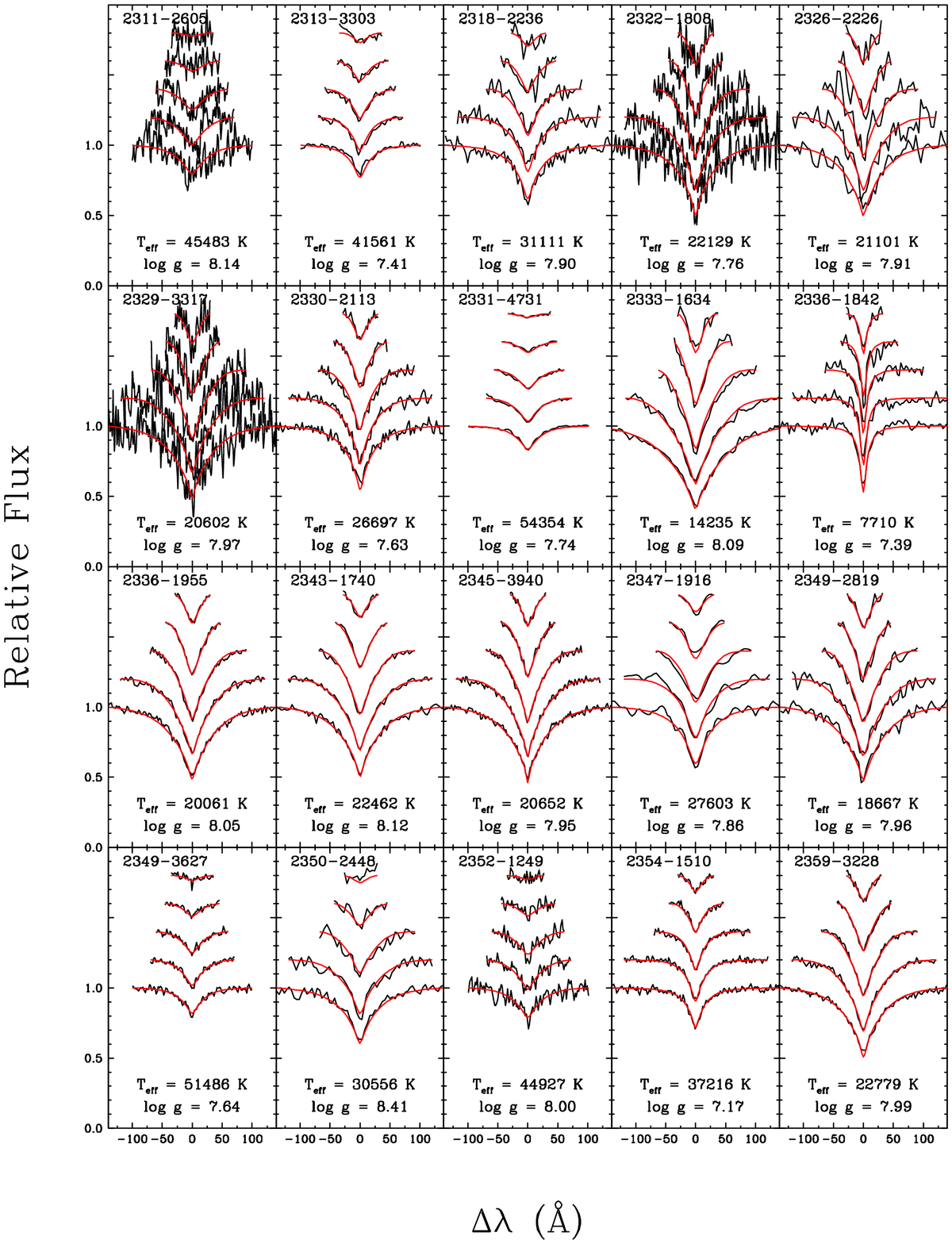}
\caption{(Continued)}
\end{figure}

\begin{figure}[p]
\centering
\includegraphics[width=0.9\linewidth]{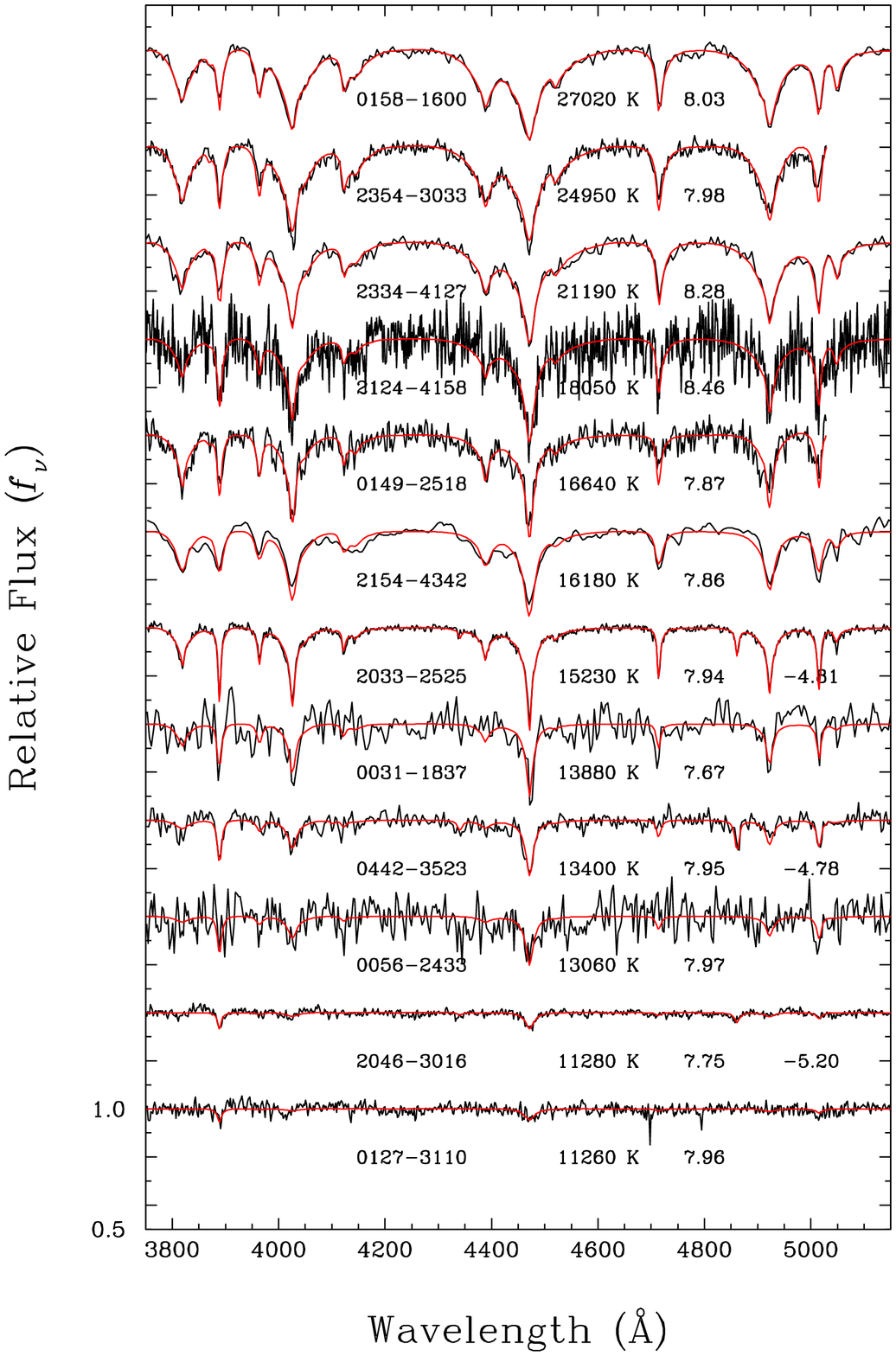}
\caption{Model fits (red) to the normalized spectra (black) for the DB and
  DBA white dwarfs in our sample, in order of decreasing effective
  temperature, from top to bottom.  The best-fit $\Te$ and
  $\logg$ values --- as well as $\log\nh$ for DBA stars --- are
  indicated under each spectrum.\label{fit_DB}}
\end{figure}

A similar approach can be used for the DB white dwarfs in our sample
following the procedure and model atmospheres described in
\citet{Bergeron2011}, with improvements to the van der Waals
broadening discussed in \citet{Genest2019}.  Here the full spectrum is
normalized to a continuum set to unity, and there is one additional
parameter to fit, the hydrogen-to-helium abundance ratio $\nh$ (in
number). Given the lack of spectral coverage in the red where \halpha\
is located, the value of $\nh$ is determined by using \hbeta, which is
detected in only three objects in our sample; we simply assume
a pure helium composition for the remaining objects. Our best fits for
the DB and DBA white dwarfs in our sample are displayed in Figure
\ref{fit_DB} in order of decreasing effective temperature. The values
of $\Te$ and $\logg$ are given in the figure, as well as $\log\nh$ in
the case of DBA stars.

\begin{figure}[p]
\centering
\includegraphics[width=0.9\linewidth]{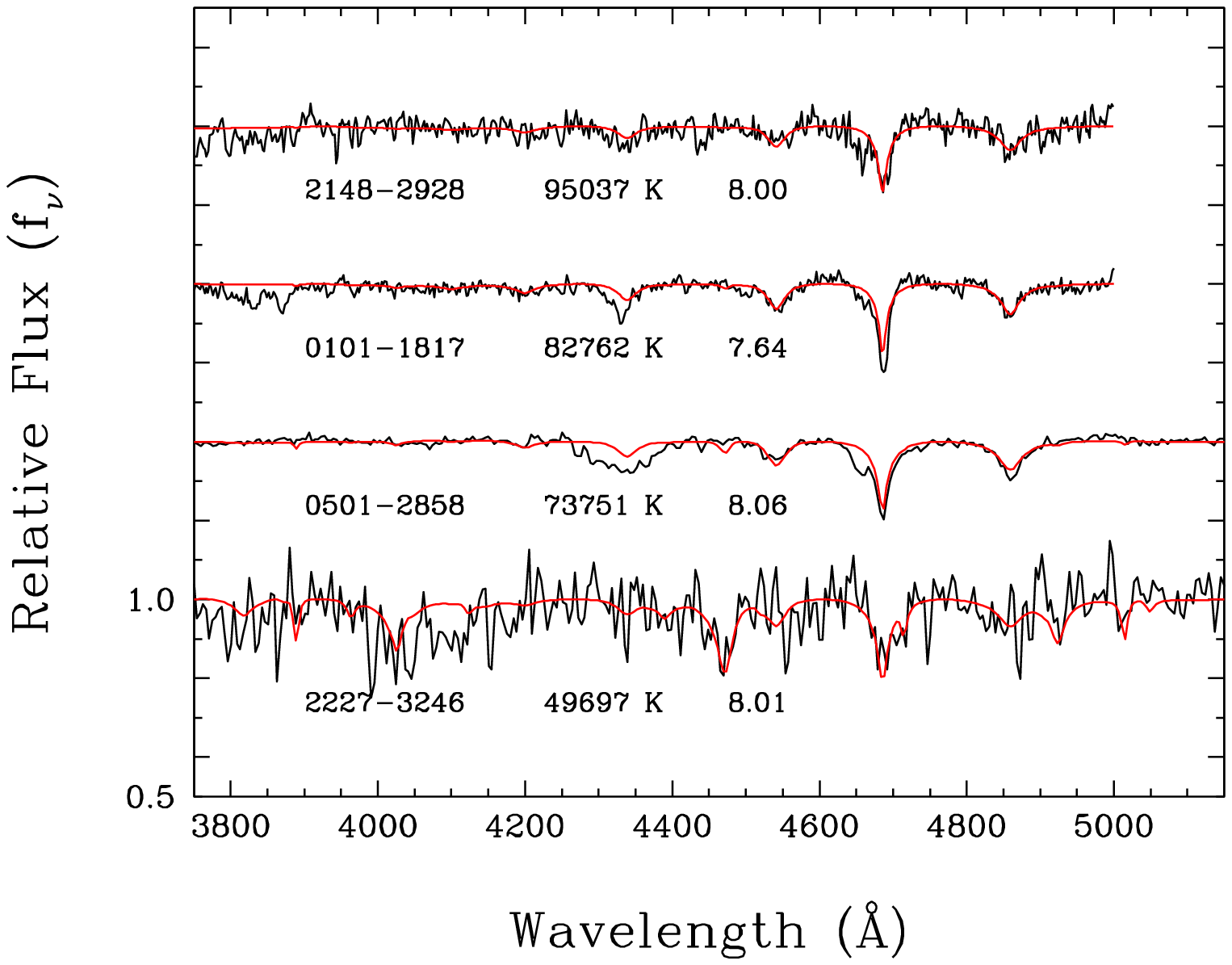}
\caption{Model fits (red) to the normalized spectra (black) for the DO
  white dwarfs in our sample, in order of decreasing effective
  temperature, from top to bottom.  The best-fit $\Te$ and
  $\logg$ values are indicated under each spectrum.\label{fit_DO}}
\end{figure}

The four DO stars in our sample are fitted using the same method and
model atmospheres as those described in detail in
\citet{Bedard2020}. A pure helium composition is assumed for all
objects. Our best fits are displayed in Figure \ref{fit_DO} in order
of decreasing effective temperature, with the values of $\Te$ and
$\logg$ given in the figure. Some of the fits displayed here show some
problems discussed at length in B\'edard et al., and we refer the
reader to this paper for a full discussion. Briefly, MCT 0101$-$1817
and MCT 0501$-$2858 show abnormally broad and deep \heii\ features
that are poorly fitted by the models, thereby revealing the presence
of a wind-fed circumstellar magnetosphere around these objects
\citep{Reindl2019,Reindl2021}. Note that MCT 0101$-$1817 also shows
so-called ultra-high excitation lines in the blue, as also pointed out
by \citet{Reindl2021}. Finally, MCT 2148$-$2928 exhibits relatively strong
\civ\ features and thus our pure-He models might not be appropriate for
this star. Consequently, our atmospheric parameters for the three
aforementioned objects should be taken with caution.

For the DC stars in our sample, it is not possible to measure the
physical parameters using the spectroscopic technique since these
objects are featureless. Here we take advantage of the available {\it
  Gaia} observational data and estimate the effective temperature and
stellar radius using the photometric technique \citep{Bergeron1997},
where the energy distribution built from the {\it Gaia} $G$, $G_{\rm
  BP}$, and $G_{\rm RP}$ photometry (EDR3) is fitted with synthetic
photometry obtained from model atmospheres. In this case, the fitted
parameters are the effective temperature, $\Te$, and the solid angle,
$\pi(R/D)^2$, where $R$ is the radius of the star, and $D$ its
distance from Earth obtained from the {\it Gaia} parallax measurement.
Also, interstellar reddening is taken into account since several
objects in our sample are sufficiently distant (see
\citealt{Bergeron2019} for details).  The photometric fits are not
shown here, and the physical parameters for these DC stars are
provided in the next section. As a reminder, we also used this
photometric approach to discriminate between the cool and hot
spectroscopic solutions for the DA stars.

\begin{figure}[p]
\centering
\includegraphics[width=0.9\linewidth]{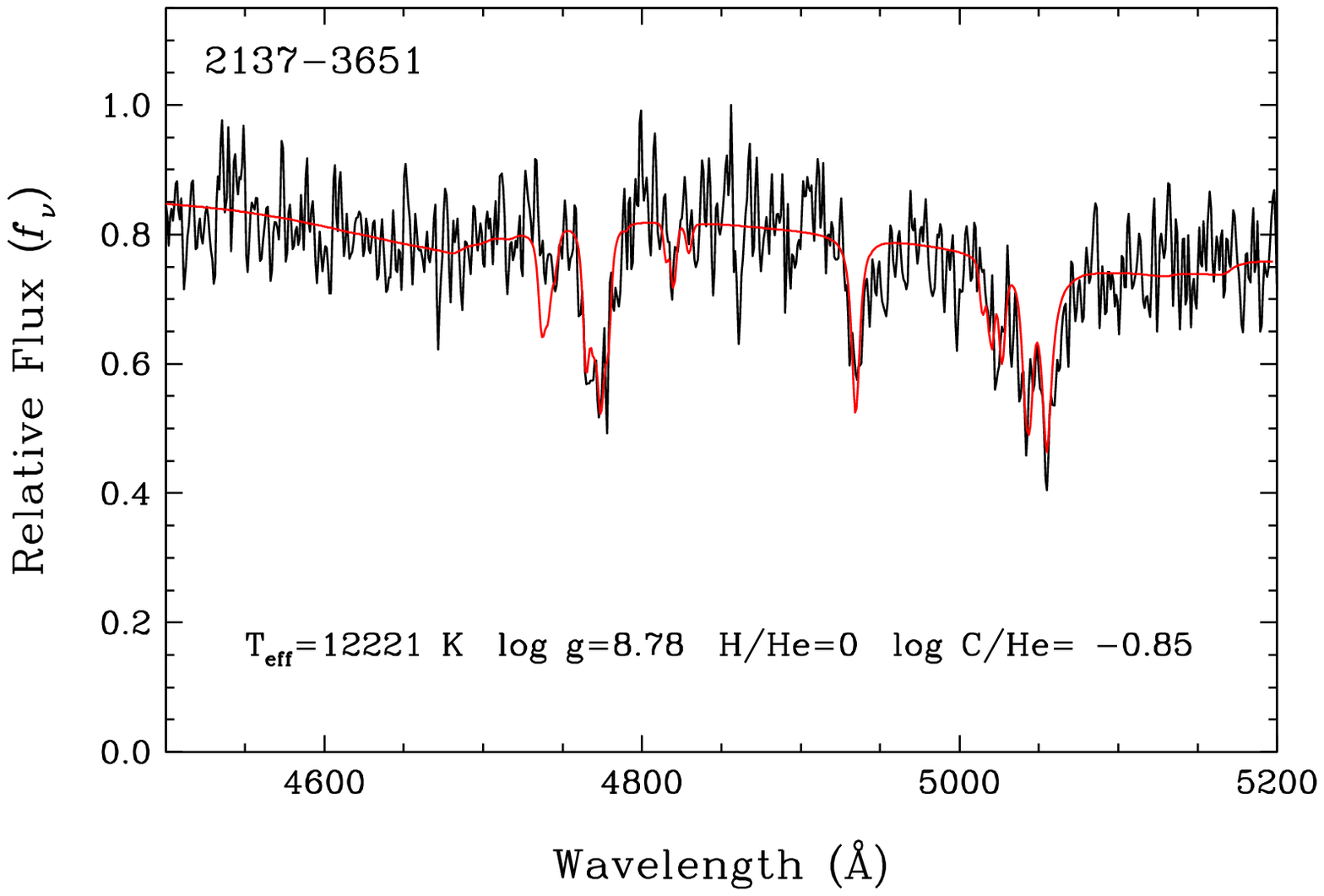}
\caption{Spectroscopic fit to the only DQ white dwarf in our
  sample. The adopted atmospheric parameters are given in the
  figure. Note that the $\logg$ value is taken from the photometric
  fit to the {\it Gaia} data, and that $\nh$ is assumed to be zero.\label{fit_DQ}}
\end{figure}

Finally, the only DQ star in our sample is fitted using a hybrid
approach between the spectroscopic and photometric methods. More
specifically, we first fit the photometric energy distribution to
measure the stellar radius and thus the mass and surface gravity
through the mass-radius relation for white dwarfs. Then we force this
$\logg$ value and fit the optical spectra with only $\Te$ and C/He ---
the carbon-to-helium abundance ratio (in number) --- as independent
parameters. We also assume $\nh=0$. Throughout, we rely on the
detailed model atmospheres for DQ stars described in
\citet{Blouin2019}.  Our best fit is displayed in Figure \ref{fit_DQ}
with the adopted atmospheric parameters reported in the figure. With
an inferred mass of 1.08 \msun\ (see below), MCT 2137$-$3651 belongs
to this sequence of massive ``warm'' DQ white dwarfs identified in
Figure 12 of \citet{Coutu2019}, with the largest carbon abundance ever
detected, although this conclusion will require a more detailed
analysis than that performed here.

\section{Global Properties of the Sample}

We summarize in Table 2 the results of our analysis for the 144 white
dwarfs in our sample, where we give for each entry the MCT number, the
spectral type, the Gaia ID (EDR3), the effective temperature ($\Te$,
rounded off to the nearest 10~K), the surface gravity ($\logg$), the
stellar mass ($M$) --- together with uncertainties
---, the assumed atmospheric composition, the parallactic distance
($D$, if available), and the white dwarf cooling time ($\log\tau$ with
$\tau$ in years). A note added in the last column indicates whether the
object has been fitted photometrically instead of spectroscopically,
and whether a spectral type is already available in the MWDD. Note
that many of these previously known spectral types are taken from the
Villanova White Dwarf Catalog of \citet{McCook1999}, where the quoted
spectral type actually comes from the MCT survey in a private
communication (see, e.g., MCT 0458$-$3020).

With both fitting techniques, one must rely on mass-radius relations
obtained from detailed evolutionary models to infer masses and cooling
ages. Here we use the improved cooling sequences and interpolation
scheme discussed at length in \citet{Bedard2020}. These models ---
also publicly available on our Web site (see footnote above) --- have
C/O-core (50/50) compositions, $q({\rm He})\equiv M_{\rm
  He}/M_{\star}=10^{-2}$, and $q({\rm H})=10^{-4}$ or $10^{-10}$,
which we use for DA and non-DA stars, respectively. Note that some of
the cooling times are omitted from Table 2 because they are too
small. As discussed by B\'edard et al., very short cooling ages ($\log
\tau\ \lta 5$) are sensitive to the zero points set by the initial
models of the cooling sequences. In such cases, one can simply state
an upper limit of $\log \tau < 5$.

As discussed above, the values of $\Te$ and $\logg$ for the DA stars
have been determined spectroscopically using model atmospheres
calculated within the mixing-length theory, the so-called 1D
models. Such 1D models are well-known to yield spurious high-$\logg$
values for white dwarfs below $\Te\sim12,000$~K (see the full
discussion of this problem in \citealt{Tremblay2010}). This
long-standing high-$\logg$ problem has been solved by
\citet{Tremblay3D2011}, who showed that the use of more sophisticated
3D hydrodynamical model atmospheres yields $\logg$ values that are
significantly reduced. Consequently, we applied in Table 2 the 3D
corrections of \citet{Tremblay2013} to both $\Te$ and $\logg$ for the
DA stars in the appropriate range of temperature where convection is
important. Note that in principle, similar 3D corrections should also
be applied to DB white dwarfs \citep{Cukanovaite2021}. However, as
discussed by Cukanovaite et al., given the uncertainties in the
microphysical calculations of DB model atmospheres, these 3D
corrections are likely to change, and we simply refrain from applying
them to the DB parameters in Table 2 (see also \citealt{Barnett2021}).

\begin{figure}[p]
\centering
\includegraphics[width=0.9\linewidth]{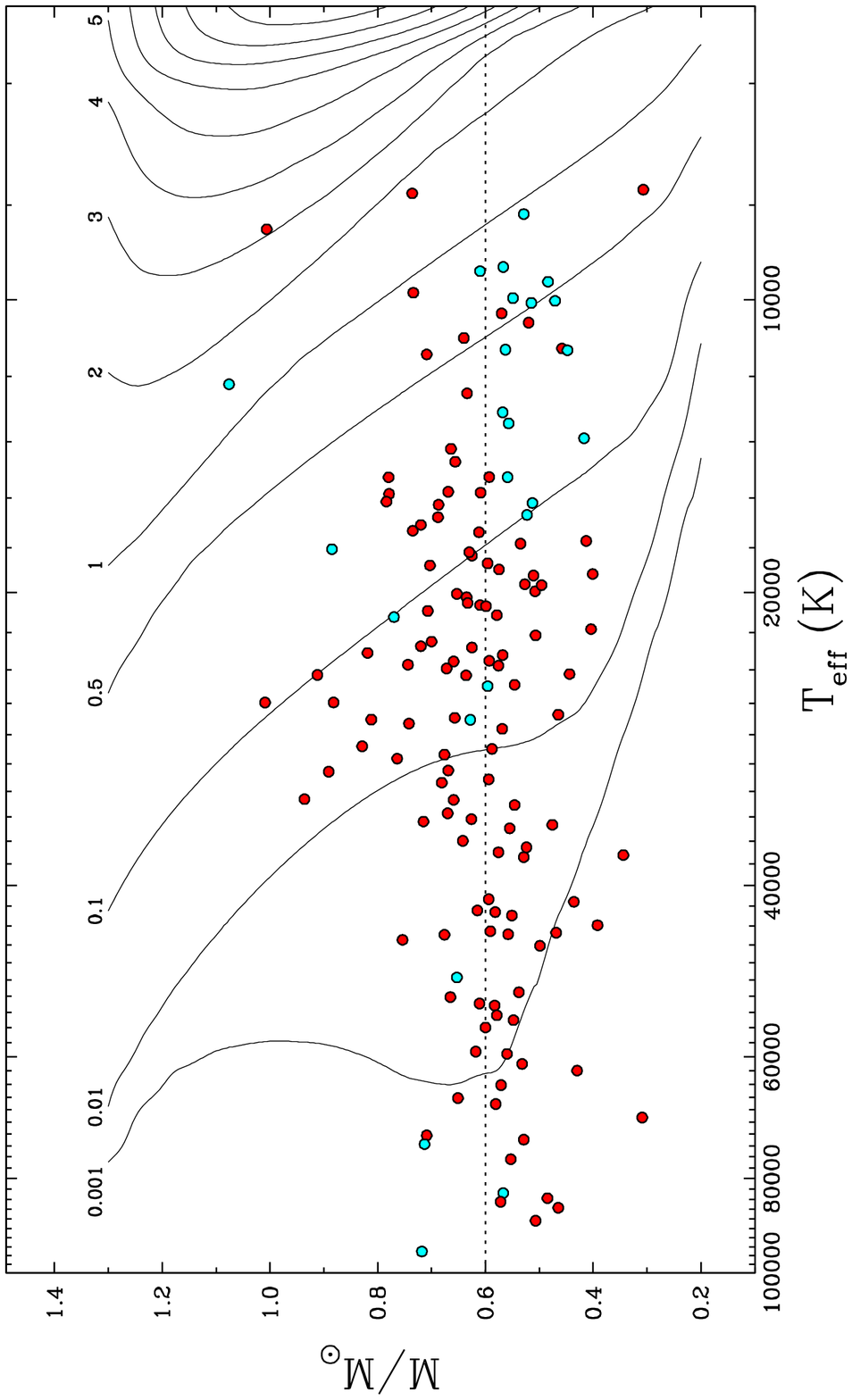}
\caption{Distribution of mass as a function of effective temperature
  for all the H-atmosphere (red symbols) and He-atmosphere (cyan
  symbols) white dwarfs in our sample. Also shown as solid lines are
  the theoretical isochrones from our CO-core evolutionary models,
  labeled with the white dwarf cooling age in Gyr. The dotted line at
  0.6 \msun\ serves as a reference.\label{correltm}}
\end{figure}

The mass distribution of all white dwarfs in the MCT sample is
displayed in Figure \ref{correltm} as a function of effective
temperature. Also shown are theoretical isochrones, labeled in units
of Gyr, obtained from our evolutionary models. In this figure, we
distinguish white dwarfs with H-dominated (red symbols) and
He-dominated (cyan symbols) atmospheres. The masses of all white
dwarfs appear evenly distributed around the mean canonical mass of 0.6
\msun, also shown in the figure. There are four known ZZ Ceti
variables in the MCT sample, indicated in Table 2; given its location
in Figure \ref{correltm}, MCT 0142$-$3026 would represent a good ZZ
Ceti candidate.

For the DC white dwarfs in our sample, which have been fitted using
the photometric technique, we assumed a varying value of $\nh$ as a
function of $\Te$, as prescribed in \citet{Kilic2020}. Such a non-zero
hydrogen abundance is required in order to avoid overestimating the
photometric masses, as discussed at length in \citet[][see their
  Figures 10 and 11]{Bergeron2019}. We note that the inferred
photometric masses for these cool DC stars are all below 0.6 \msun,
but that they also align nicely with the warmer DB white dwarfs
analyzed with the spectroscopic technique.

Our white dwarf sample includes the usual high- and low-mass objects
usually identified in such surveys. In particular, MCT 2035$-$4236,
MCT 2336$-$1842, and MCT 2354$-$1510 have spectroscopic masses well
below 0.4 \msun, and are thus most likely unresolved double degenerate
binaries. We also include in this list of double degenerate candidates
(noted in Table 2) two white dwarfs with extremely low photometric
masses, and with photometric temperatures that differ significantly
from the spectroscopic values: MCT 0028$-$4729 ($T_{\rm
  phot}=13,148$~K, $M_{\rm phot}=0.234$ \msun) and MCT 0145$-$7035
($T_{\rm phot}=12,813$~K, $M_{\rm phot}=0.289$ \msun).

\begin{figure}[p]
\centering
\includegraphics[width=0.9\linewidth]{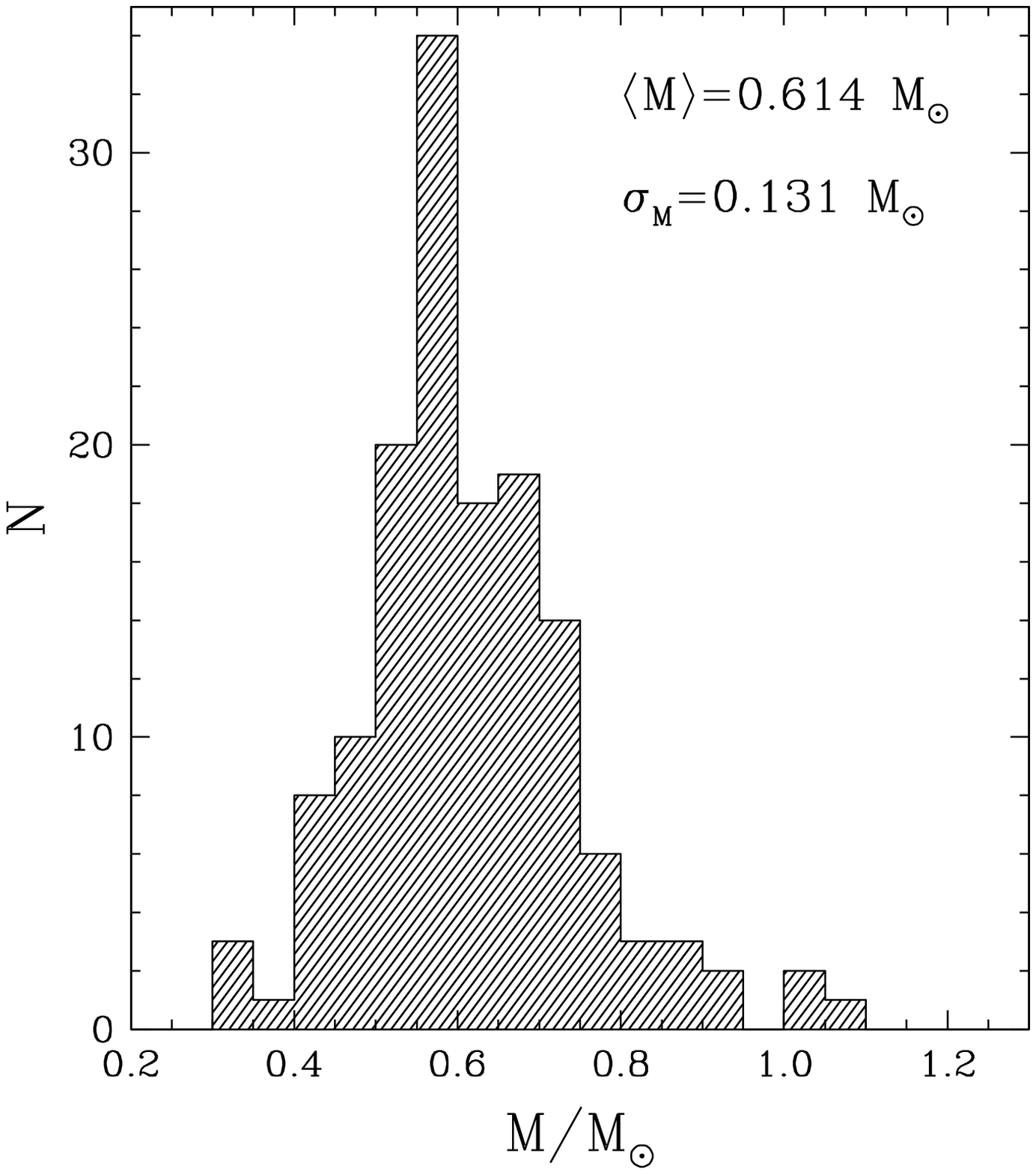}
\caption{Cumulative mass distribution for all white dwarfs in our
  sample. The mean mass and standard deviation are given in the
  figure.\label{histo}}
\end{figure}

The cumulative mass distribution, regardless of temperature, is
displayed in Figure \ref{histo}. The mean mass of our sample, $\langle
M\rangle=0.614$ \msun, and dispersion, $\sigma_M=0.131$ \msun, are
entirely consistent with the values reported by \citet[][see their
  Figure 21]{Genest2019} for the DA and DB white dwarfs in the SDSS.
Only three white dwarfs with masses in excess of 1.0 \msun\ can be
identified in Figures \ref{correltm} and \ref{histo}: MCT
0136$-$2010, MCT 0308$-$2305, and MCT 2137$-$3651 (the only DQ star in
our sample). This is typical of UV-excess magnitude-limited surveys, which tend
to underestimate the number of massive white dwarfs, due to their
intrinsic small radii and low luminosities \citep{Liebert2005}.

\section{Concluding Remarks}

We presented spectroscopic observations of 144 white dwarfs secured in
the course of the MCT survey in the southern hemisphere, including 120
DA, 12 DB, 4 DO, 1 DQ, and 7 DC stars. Our main goal was to provide
spectral types for the ongoing effort to confirm spectroscopically all
white dwarf candidates in the {\it Gaia} survey, as well as to make
available our spectra to the scientific community through the MWDD. We
also included in the MWDD the spectra of MCT 0128$-$3846, MCT 0130$-$1937, and MCT
0453$-$2933, displayed in Figure \ref{spec_other} and already
analyzed elsewhere in the literature.  Although not discussed here,
the spectra of many hot subdwarfs have been secured as part of the MCT
survey, and these can be made available upon request.

\acknowledgements We would like to dedicate this paper to two of the
coauthors, F.~Wesemael and G.~Fontaine, who passed away in recent
years (2011 and 2019, respectively). Their contribution to the white
dwarf field will long be remembered. We are grateful to the CTIO Time
Allocation Committee for its unswerving support of this project, and
to the CTIO staff for its help and technical support over the
years. L.~Murphy and A.~Beauchamp participated to some of the
observing runs at Cerro Tololo and Las Campanas. This work was
supported in part by the NSERC Canada, by the Fund FQR-NT (Qu\'ebec),
and by NATO.

\clearpage
\clearpage
\begin{deluxetable}{lccccclc}
\tabletypesize{\scriptsize}
\tablecolumns{8}
\tablewidth{0pt}
\tablecaption{Photometric and Spectroscopic Data}
\tablehead{
\colhead{MCT} &
\colhead{RA (B1950.0)} &
\colhead{Dec (B1950.0)} &
\colhead{$B_{\rm pg}$} &
\colhead{$(U-B)_{\rm pg}$} &
\colhead{Date} &
\colhead{Telescope} &
\colhead{Res (\AA)}}
\startdata
0000$-$1838&00 00 37.71& $-$18 38 39.1&16.29& $-$0.82&16/10/89&CTIO 4 m      & 5\\
0012$-$1720&00 12 45.41& $-$17 20 45.0&16.75& $-$1.02&15/10/89&CTIO 4 m      & 5\\
0016$-$2553&00 16 13.26& $-$25 53 19.1&16.07& $-$0.74&09/10/89&CTIO 4 m      & 3\\
0024$-$1211&00 24 00.70& $-$12 11 23.2&16.26& $-$0.66&16/10/89&CTIO 4 m      & 5\\
0027$-$6341&00 27 39.71& $-$63 41 31.4&15.29& $-$1.03&13/11/92&CTIO 1.5 m    & 6\\
0028$-$4729&00 28 22.78& $-$47 29 10.0&15.34& $-$0.69&02/10/88&CTIO 1.5 m    & 6\\
0031$-$1837&00 31 10.35& $-$18 37 01.5&15.68& $-$0.88&08/10/89&CTIO 1.5 m    & 6\\
0031$-$3107&00 31 56.97& $-$31 07 56.7&16.43& $-$0.96&06/12/95&CTIO 1.5 m    & 5\\
0032$-$1313&00 32 55.84& $-$13 13 36.1&16.82& $-$0.93&08/12/95&CTIO 1.5 m    & 5\\
0032$-$1735&00 32 46.26& $-$17 35 21.6&15.38& $-$0.75&06/10/89&CTIO 1.5 m    & 6\\
0032$-$3146&00 32 22.17& $-$31 46 26.5&15.62& $-$0.91&08/10/89&CTIO 1.5 m    & 6\\
0033$-$3440&00 33 47.03& $-$34 40 01.8&16.42& $-$0.69&06/12/95&CTIO 1.5 m    & 5\\
0049$-$2658&00 49 08.82& $-$26 58 13.9&16.42& $-$0.91&07/12/95&CTIO 1.5 m    & 5\\
0050$-$3316&00 50 53.50& $-$33 16 12.3&13.30& $-$1.20&12/11/92&CTIO 1.5 m    & 6\\
0052$-$1442&00 52 26.04& $-$14 42 20.3&14.95& $-$1.03&04/09/85&CTIO 1.5 m    &13\\
0056$-$2433&00 56 43.45& $-$24 33 40.7&16.36& $-$0.73&07/12/95&CTIO 1.5 m    & 5\\
0101$-$1817&01 01 47.32& $-$18 17 45.3&15.26& $-$1.00&04/09/85&CTIO 1.5 m    &13\\
0102$-$1835&01 02 25.55& $-$18 35 48.7&16.40& $-$0.61&14/09/94&CTIO 4 m      & 3\\
0105$-$1634&01 05 41.73& $-$16 34 39.4&16.44& $-$0.73&14/09/94&CTIO 4 m      & 3\\
0106$-$3550&01 06 01.02& $-$35 50 39.1&14.75& $-$1.08&04/09/85&CTIO 1.5 m    &13\\
0110$-$1355&01 10 41.53& $-$13 55 19.6&15.68& $-$0.96&07/10/89&CTIO 1.5 m    & 6\\
0110$-$1617&01 10 46.37& $-$16 17 36.7&16.37& $-$0.60&07/12/95&CTIO 1.5 m    & 5\\
0111$-$3806&01 11 45.83& $-$38 06 33.5&15.48& $-$1.09&08/10/89&CTIO 1.5 m    & 6\\
0112$-$1931&01 12 39.31& $-$19 31 07.0&16.20& $-$0.73&15/10/89&CTIO 4 m      & 5\\
0118$-$1638&01 18 31.78& $-$16 38 28.9&15.90& $-$0.64&09/10/89&CTIO 1.5 m    & 6\\
0124$-$2546&01 24 33.91& $-$25 46 20.5&15.66& $-$0.60&09/10/89&CTIO 1.5 m    & 6\\
0127$-$3110&01 27 37.56& $-$31 10 37.6&15.61& $-$0.81&08/10/89&CTIO 1.5 m    & 6\\
0129$-$2035&01 29 14.90& $-$20 35 22.8&14.64& $-$0.62&01/10/88&CTIO 1.5 m    & 6\\
0130$-$6846&01 30 30.27& $-$68 46 35.2&14.53& $-$0.98&16/11/86&CTIO 1.5 m    & 7\\
0131$-$1622&01 31 58.00& $-$16 22 26.2&13.96& $-$1.38&31/08/85&CTIO 1.5 m    &13\\
0133$-$3846&01 33 03.94& $-$38 46 52.0&16.32& $-$0.69&14/10/89&CTIO 4 m      & 5\\
0134$-$4042&01 34 37.67& $-$40 42 49.4&16.46& $-$0.86&14/09/94&CTIO 4 m      & 3\\
0136$-$2010&01 36 08.17& $-$20 10 01.5&16.49& $-$0.64&01/07/94&La Palma 2.5 m& 6\\
0137$-$2908&01 37 58.49& $-$29 08 01.8&16.04& $-$0.65&09/10/89&CTIO 4 m      & 3\\
0138$-$2336&01 38 06.61& $-$23 36 29.5&16.10& $-$0.89&09/10/89&CTIO 4 m      & 3\\
0138$-$4014&01 38 00.87& $-$40 14 34.9&16.10& $-$0.67&15/09/94&CTIO 4 m      & 3\\
0140$-$3914&01 40 40.47& $-$39 14 12.7&14.43& $-$0.97&14/11/86&CTIO 1.5 m    & 7\\
0142$-$3026&01 42 19.45& $-$30 26 24.5&16.46& $-$0.65&08/12/95&CTIO 1.5 m    & 5\\
0145$-$2211&01 45 00.17& $-$22 11 46.6&15.30& $-$0.66&29/09/88&CTIO 1.5 m    & 6\\
0145$-$7035&01 45 01.40& $-$70 35 15.6&15.31& $-$0.82&02/10/88&CTIO 1.5 m    & 6\\
0146$-$2500&01 46 19.70& $-$25 00 37.4&16.37& $-$0.69&08/12/95&CTIO 1.5 m    & 5\\
0149$-$2518&01 49 40.50& $-$25 18 00.4&15.91& $-$0.72&14/10/89&CTIO 4 m      & 5\\
0154$-$2405&01 54 52.62& $-$24 05 59.0&16.36& $-$1.11&24/09/88&CTIO 4.0 m    & 7\\
0158$-$1600&01 58 32.36& $-$16 00 38.9&14.38& $-$0.99&29/09/88&CTIO 1.5 m    & 6\\
0158$-$2242&01 58 33.57& $-$22 42 03.5&16.10& $-$1.01&15/10/89&CTIO 4 m      & 5\\
0200$-$1243&02 00 38.15& $-$12 43 21.8&14.75& $-$0.89&06/10/89&CTIO 1.5 m    & 6\\
0203$-$1807&02 03 01.31& $-$18 07 43.8&15.48& $-$0.69&06/10/89&CTIO 1.5 m    & 6\\
0205$-$3025&02 05 27.35& $-$30 25 09.8&15.67& $-$0.60&01/10/88&CTIO 1.5 m    & 6\\
0205$-$3635&02 05 17.64& $-$36 35 03.1&16.10& $-$1.00&15/10/89&CTIO 4 m      & 5\\
0208$-$1520&02 08 18.90& $-$15 20 40.4&15.56& $-$0.67&07/10/89&CTIO 1.5 m    & 6\\
0208$-$2621&02 08 42.66& $-$26 21 06.7&16.15& $-$0.83&15/10/89&CTIO 4 m      & 5\\
0210$-$3929&02 10 39.93& $-$39 29 54.4&16.48& $-$0.64&08/12/95&CTIO 1.5 m    & 5\\
0212$-$2308&02 12 03.16& $-$23 08 46.5&16.08& $-$0.73&15/10/89&CTIO 4 m      & 5\\
0219$-$4049&02 19 18.92& $-$40 49 10.7&16.21& $-$0.66&05/12/95&CTIO 1.5 m    & 5\\
0221$-$2642&02 21 15.04& $-$26 42 54.0&15.62& $-$1.21&11/09/94&CTIO 1.5 m    & 6\\
0222$-$2630&02 22 21.60& $-$26 30 24.1&15.55& $-$0.77&11/09/94&CTIO 1.5 m    & 6\\
0226$-$3255&02 26 19.39& $-$32 55 50.7&13.90& $-$1.03&13/11/86&CTIO 1.5 m    & 7\\
0235$-$1234&02 35 01.60& $-$12 34 31.3&14.78& $-$1.11&13/11/92&CTIO 1.5 m    & 6\\
0252$-$3501&02 52 35.43& $-$35 01 58.3&15.80& $-$0.63&17/10/89&CTIO 4 m      & 5\\
0257$-$3339&02 57 01.69& $-$33 39 31.5&16.03& $-$1.02&04/12/95&CTIO 1.5 m    & 5\\
0300$-$2313&03 00 23.21& $-$23 13 35.5&15.66& $-$1.02&12/09/94&CTIO 1.5 m    & 6\\
0308$-$2305&03 08 54.29& $-$23 05 21.3&15.19& $-$1.03&12/09/94&CTIO 1.5 m    & 6\\
0309$-$2105&03 09 43.30& $-$21 05 22.5&15.76& $-$0.81&06/12/95&CTIO 1.5 m    & 5\\
0309$-$2730&03 09 25.05& $-$27 30 41.1&16.21& $-$0.80&24/09/88&CTIO 4.0 m    & 7\\
0315$-$3313&03 15 25.40& $-$33 13 57.8&16.53& $-$0.97&23/09/88&CTIO 4.0 m    & 7\\
0320$-$5355&03 20 51.40& $-$53 55 55.3&14.99& $-$1.19&25/09/93&CTIO 1.5 m    & 6\\
0333$-$3500&03 33 38.11& $-$35 00 07.5&15.51& $-$0.77&13/11/86&CTIO 1.5 m    & 7\\
0335$-$4205&03 35 29.96& $-$42 05 06.7&15.98& $-$1.41&06/12/95&CTIO 1.5 m    & 5\\
0341$-$6719&03 41 59.62& $-$67 19 03.9&15.78& $-$0.82&07/12/95&CTIO 1.5 m    & 5\\
0400$-$4045&04 00 57.85& $-$40 45 59.8&15.75& $-$0.67&06/12/95&CTIO 1.5 m    & 5\\
0413$-$4029&04 13 57.09& $-$40 29 56.3&15.98& $-$0.96&07/12/95&CTIO 1.5 m    & 5\\
0420$-$7310&04 20 23.65& $-$73 10 45.0&15.49& $-$0.72&30/09/88&CTIO 1.5 m    & 6\\
0442$-$3523&04 42 38.30& $-$35 23 14.0&15.72& $-$0.68&07/12/95&CTIO 1.5 m    & 5\\
0455$-$2812&04 55 13.97& $-$28 12 25.3&13.91& $-$1.28&00/00/00&Steward 2.3 m & 6\\
0458$-$3020&04 58 46.54& $-$30 20 56.1&16.30& $-$0.99&17/10/89&CTIO 4 m      & 5\\
0501$-$2858&05 01 57.61& $-$28 58 39.1&14.09& $-$1.36&13/11/86&CTIO 1.5 m    & 7\\
1950$-$4314&19 50 18.33& $-$43 14 59.7&14.89& $-$1.41&31/08/85&CTIO 1.5 m    &13\\
1958$-$5008&19 58 04.38& $-$50 08 23.2&15.17& $-$1.09&26/07/89&MWLCO 2.5 m   & 3\\
2000$-$5611&20 00 18.67& $-$56 11 15.4&14.97& $-$1.27&27/07/89&MWLCO 2.5 m   & 3\\
2001$-$5349&20 01 03.48& $-$53 49 22.9&16.22& $-$0.67&14/10/89&CTIO 4 m      & 5\\
2002$-$5405&20 02 33.27& $-$54 05 42.2&16.14& $-$1.20&14/10/89&CTIO 4 m      & 5\\
2004$-$3021&20 04 25.00& $-$30 21 08.4&16.03& $-$0.69&08/10/89&CTIO 1.5 m    & 6\\
2017$-$3726&20 17 00.55& $-$37 26 09.8&16.43& $-$0.79&08/08/88&MWLCO 2.5 m   & 3\\
2019$-$3438&20 19 45.79& $-$34 38 11.7&16.46& $-$1.04&08/08/88&MWLCO 2.5 m   & 3\\
2020$-$4234&20 20 35.96& $-$42 34 08.8&14.79& $-$0.81&10/08/88&MWLCO 2.5 m   & 3\\
2031$-$2745&20 31 53.30& $-$27 45 10.6&14.73& $-$1.16&11/09/94&CTIO 1.5 m    & 6\\
2033$-$2525&20 33 47.24& $-$25 25 06.5&15.12& $-$1.02&11/09/94&CTIO 1.5 m    & 6\\
2035$-$3337&20 35 30.02& $-$33 37 09.6&14.57& $-$1.02&10/08/88&MWLCO 2.5 m   & 3\\
2035$-$4236&20 35 48.09& $-$42 36 46.9&16.86& $-$1.30&09/08/88&MWLCO 2.5 m   & 3\\
2040$-$3914&20 40 34.06& $-$39 14 05.4&14.29& $-$0.64&10/08/88&MWLCO 2.5 m   & 3\\
2046$-$3016&20 46 28.09& $-$30 16 10.9&15.74& $-$0.88&26/07/89&MWLCO 2.5 m   & 3\\
2100$-$3821&21 00 03.78& $-$38 21 33.6&15.97& $-$0.84&09/08/88&MWLCO 2.5 m   & 3\\
2100$-$3840&21 00 11.66& $-$38 40 09.3&16.23& $-$0.96&09/08/88&MWLCO 2.5 m   & 3\\
2101$-$3627&21 01 38.84& $-$36 27 22.5&16.10& $-$0.91&10/08/88&MWLCO 2.5 m   & 3\\
2103$-$3947&21 03 19.75& $-$39 47 55.9&15.42& $-$0.48&28/09/88&CTIO 1.5 m    & 6\\
2107$-$2653&21 07 18.35& $-$26 53 39.2&16.11& $-$1.10&14/10/89&CTIO 4 m      & 5\\
2108$-$4310&21 08 21.45& $-$43 10 27.3&16.03& $-$0.67&15/10/89&CTIO 4 m      & 5\\
2109$-$2934&21 09 41.30& $-$29 34 36.2&15.44& $-$0.65&26/07/89&MWLCO 2.5 m   & 3\\
2113$-$3705&21 13 41.55& $-$37 05 05.8&16.38& $-$0.67&11/08/88&MWLCO 2.5 m   & 3\\
2114$-$3737&21 14 17.52& $-$37 37 07.2&15.64& $-$0.76&08/08/88&MWLCO 2.5 m   & 3\\
2122$-$4643&21 22 11.83& $-$46 43 33.9&15.99& $-$0.67&15/10/89&CTIO 4 m      & 5\\
2124$-$4158&21 24 25.23& $-$41 58 22.1&15.89& $-$0.89&13/08/88&MWLCO 2.5 m   & 3\\
2133$-$3637&21 33 35.91& $-$36 37 26.8&15.74& $-$0.80&08/08/88&MWLCO 2.5 m   & 3\\
2137$-$3651&21 37 21.65& $-$36 51 33.4&15.96& $-$0.85&10/08/88&MWLCO 2.5 m   & 3\\
2137$-$3756&21 37 14.63& $-$37 56 21.6&16.03& $-$0.67&11/08/88&MWLCO 2.5 m   & 3\\
2146$-$4320&21 46 31.05& $-$43 20 12.8&15.85& $-$1.21&10/08/88&MWLCO 2.5 m   & 3\\
2148$-$2910&21 48 45.90& $-$29 10 51.7&15.99& $-$0.64&14/10/89&CTIO 4 m      & 5\\
2148$-$2928&21 48 24.79& $-$29 28 43.4&16.30& $-$0.98&15/10/89&CTIO 4 m      & 5\\
2151$-$3043&21 51 58.72& $-$30 43 31.9&14.82& $-$0.78&26/07/89&MWLCO 2.5 m   & 3\\
2153$-$4156&21 53 30.56& $-$41 56 31.2&15.67& $-$1.05&11/08/88&MWLCO 2.5 m   & 3\\
2154$-$4342&21 54 54.79& $-$43 42 12.9&14.82& $-$0.83&02/09/85&CTIO 1.5 m    &13\\
2159$-$4129&21 59 25.55& $-$41 29 00.6&15.54& $-$1.09&03/09/85&CTIO 1.5 m    &13\\
2214$-$3740&22 14 49.76& $-$37 40 42.3&15.93& $-$0.99&01/10/88&CTIO 1.5 m    & 6\\
2227$-$3246&22 27 58.77& $-$32 46 56.3&15.84& $-$0.97&01/10/88&CTIO 1.5 m    & 6\\
2231$-$2935&22 31 39.20& $-$29 35 15.1&15.73& $-$0.98&26/07/89&MWLCO 2.5 m   & 3\\
2233$-$3529&22 33 58.95& $-$35 29 40.9&15.28& $-$0.89&10/09/94&CTIO 1.5 m    & 6\\
2241$-$4031&22 41 28.80& $-$40 31 21.7&15.99& $-$0.79&14/09/94&CTIO 4 m      & 3\\
2251$-$6326&22 51 58.89& $-$63 26 30.0&14.28& $-$0.96&13/08/88&MWLCO 2.5 m   & 3\\
2252$-$4042&22 52 51.57& $-$40 42 21.0&15.70& $-$1.09&12/09/94&CTIO 1.5 m    & 6\\
2259$-$2646&22 59 20.34& $-$26 46 56.5&15.18& $-$0.76&13/08/88&MWLCO 2.5 m   & 3\\
2259$-$3212&22 59 28.62& $-$32 12 07.4&15.98& $-$1.07&23/09/88&CTIO 4.0 m    & 7\\
2306$-$2726&23 06 10.54& $-$27 26 41.9&15.88& $-$0.56&30/09/88&CTIO 1.5 m    & 6\\
2311$-$2605&23 11 12.69& $-$26 05 04.5&16.05& $-$1.23&12/08/88&MWLCO 2.5 m   & 3\\
2313$-$3303&23 13 20.65& $-$33 03 02.2&15.74& $-$1.09&23/09/88&CTIO 4.0 m    & 7\\
2318$-$2236&23 18 47.63& $-$22 36 40.9&16.05& $-$0.88&23/09/88&CTIO 4.0 m    & 7\\
2322$-$1808&23 22 41.04& $-$18 08 27.1&15.31& $-$0.96&11/08/88&MWLCO 2.5 m   & 3\\
2326$-$2226&23 26 01.11& $-$22 26 48.0&15.85& $-$0.48&30/09/88&CTIO 1.5 m    & 6\\
2329$-$3317&23 29 31.79& $-$33 17 37.2&16.24& $-$0.78&12/08/88&MWLCO 2.5 m   & 3\\
2330$-$2113&23 30 22.59& $-$21 13 44.1&16.25& $-$0.74&14/10/89&CTIO 4 m      & 5\\
2331$-$4731&23 31 19.78& $-$47 31 00.1&13.59& $-$1.18&18/08/87&CTIO 1.5 m    & 6\\
2333$-$1634&23 33 00.57& $-$16 34 16.3&13.80& $-$0.91&28/09/88&CTIO 1.5 m    & 6\\
2334$-$4127&23 34 58.73& $-$41 27 07.7&15.28& $-$1.15&09/09/94&CTIO 1.5 m    & 6\\
2336$-$1842&23 36 16.75& $-$18 42 46.9&15.84& $-$0.64&23/09/88&CTIO 4.0 m    & 7\\
2336$-$1955&23 36 51.69& $-$19 55 23.4&16.30& $-$0.74&16/10/89&CTIO 4 m      & 5\\
2343$-$1740&23 43 50.50& $-$17 40 54.1&16.12& $-$1.04&01/07/94&La Palma 2.5 m& 6\\
2345$-$3940&23 45 49.42& $-$39 40 27.2&16.06& $-$1.01&14/09/94&CTIO 4 m      & 3\\
2347$-$1916&23 47 28.08& $-$19 16 02.0&15.38& $-$0.90&03/09/85&CTIO 1.5 m    &13\\
2349$-$2819&23 49 47.87& $-$28 19 53.9&15.54& $-$0.86&06/10/89&CTIO 1.5 m    & 6\\
2349$-$3627&23 49 32.40& $-$36 27 22.9&16.41& $-$1.18&14/09/94&CTIO 4 m      & 3\\
2350$-$2448&23 50 28.94& $-$24 48 43.4&15.21& $-$1.02&29/09/88&CTIO 1.5 m    & 6\\
2352$-$1249&23 52 39.63& $-$12 49 34.4&16.49& $-$0.71&15/09/94&CTIO 4 m      & 3\\
2354$-$1510&23 54 59.41& $-$15 10 49.0&15.02& $-$0.98&01/10/88&CTIO 1.5 m    & 6\\
2354$-$3033&23 54 02.78& $-$30 33 02.1&16.26& $-$0.78&16/10/89&CTIO 4 m      & 5\\
2359$-$3228&23 59 58.46& $-$32 28 26.1&15.87& $-$0.64&24/09/88&CTIO 4.0 m    & 7\\
\enddata
\end{deluxetable}

\clearpage
\clearpage
\begin{deluxetable}{llcrrlccccccc}
\tabletypesize{\scriptsize}
\tablecolumns{10}
\tablewidth{0pt}
\tablecaption{Atmospheric Parameters of MCT White Dwarfs}
\tablehead{
\colhead{MCT} &
\colhead{Type} &
\colhead{Gaia ID (EDR3)} &
\colhead{$\Te$ (K)} &
\colhead{log $g$} &
\colhead{$M/$\msun} &
\colhead{Comp} &
\colhead{$D ({\rm pc})$} &
\colhead{log $\tau$} &
\colhead{Notes}}
\startdata
0000$-$1838&   DA     &2414099622710507904 &15,220 (279)        &7.96 (0.05)    &0.59 (0.03)    &      H        &   99&8.26      &1                   \\
0012$-$1720&   DA     &2368091860020553472 &54,980 (1809)       &7.64 (0.12)    &0.55 (0.04)    &      H        &  360&6.09      &                    \\
0016$-$2553&   DA     &2323704339384503040 &10,950 (176)        &8.06 (0.06)    &0.64 (0.04)    &      H        &   61&8.74      &1, 2                \\
0024$-$1211&   DA     &2423844216310164864 &15,790 (285)        &7.99 (0.05)    &0.61 (0.03)    &      H        &  104&8.23      &1                   \\
0027$-$6341&   DA     &4900807999725863296 &59,560 (1615)       &7.64 (0.10)    &0.56 (0.03)    &      H        &  197&5.99      &1                   \\
\\
0028$-$4729&   DA     &4978793541987799040 &17,820 (357)        &7.84 (0.06)    &0.54 (0.03)    &      H        &   96&7.89      &3                   \\
0031$-$1837&   DB     &2363965672055370112 &13,880 (716)        &7.67 (0.43)    &0.42 (0.19)    &      He       &  113&8.19      &1                   \\
0031$-$3107&   DA     &2317553529604662016 &42,420 (1474)       &7.88 (0.15)    &0.61 (0.07)    &      H        &  439&6.50      &1                   \\
0032$-$1313&   DA+dM  &2375355375568178304 &72,230 (6089)       &7.96 (0.30)    &0.71 (0.13)    &      H        &  263&5.84      &                    \\
0032$-$1735&   DA     &2364319061964016512 & 9840 (180)         &8.22 (0.12)    &0.73 (0.07)    &      H        &   33&8.95      &1                   \\
\\
0032$-$3146&   DA     &2317319612801004416 &43,930 (994)        &7.26 (0.09)    &0.39 (0.03)    &      H        &  431&---       &1                   \\
0033$-$3440&   DA     &5005361213245945856 &15,230 (515)        &8.27 (0.07)    &0.78 (0.05)    &      H        &   87&8.48      &1                   \\
0049$-$2658&   DA     &2342910436701710080 &34,380 (837)        &8.11 (0.14)    &0.71 (0.08)    &      H        &  277&6.77      &1                   \\
0050$-$3316&   DA     &5006486048001153792 &35,980 (575)        &7.97 (0.06)    &0.64 (0.03)    &      H        &   59&6.70      &1                   \\
0052$-$1442&   DA     &2372473658671416960 &25,940 (485)        &8.41 (0.07)    &0.88 (0.04)    &      H        &   73&7.82      &1                   \\
\\
0056$-$2433&   DB     &2345088259997529856 &13,060 (861)        &7.97 (0.56)    &0.57 (0.30)    &      He       &  102&8.47      &                    \\
0101$-$1817&   DOZ    &2358027755213297408 &82,760 (4624)       &7.64 (0.20)    &0.57 (0.07)    &      He       &  400&5.45      &1                   \\
0102$-$1835&   DA     &2357946464368276224 &23,780 (381)        &7.89 (0.05)    &0.58 (0.03)    &      H        &  212&7.33      &1                   \\
0105$-$1634&   DA     &2359382697136585728 &28,950 (436)        &7.89 (0.05)    &0.59 (0.03)    &      H        &  260&7.00      &1                   \\
0106$-$3550&   DA     &5014009353235167744 &30,470 (501)        &8.04 (0.07)    &0.67 (0.04)    &      H        &   93&6.95      &1                   \\
\\
0110$-$1355&   DA     &2456122476087944960 &26,900 (800)        &8.03 (0.12)    &0.66 (0.07)    &      H        &  121&7.18      &1                   \\
0110$-$1617&   DA     &2358739727648116608 &37,400 (1324)       &7.72 (0.17)    &0.53 (0.07)    &      H        &  345&6.63      &1                   \\
0111$-$3806&   DA     &4988781711771040128 &83,790 (2744)       &7.12 (0.09)    &0.48 (0.02)    &      H        &  488&---       &1                   \\
0112$-$1931&   DA     &2354521726165844096 &36,970 (899)        &7.83 (0.11)    &0.58 (0.05)    &      H        &  307&6.67      &1                   \\
0118$-$1638&   DA     &2454507602744210432 &14,680 (510)        &8.07 (0.08)    &0.66 (0.05)    &      H        &   77&8.39      &1                   \\
\\
0124$-$2546&   DA     &5037128131397095168 &23,510 (541)        &7.93 (0.07)    &0.59 (0.04)    &      H        &  162&7.38      &1                   \\
0127$-$3110&   DB     &5016897564123557248 &11,260 (479)        &7.96 (0.40)    &0.56 (0.22)    &      He       &   55&8.66      &1                   \\
0129$-$2035&   DA     &5043926824108837376 &20,230 (424)        &8.02 (0.06)    &0.64 (0.04)    &      H        &   84&7.82      &1                   \\
0130$-$6846&   DA     &4691571967756922368 &34,650 (625)        &7.60 (0.09)    &0.48 (0.03)    &      H        &  210&6.64      &1                   \\
0131$-$1622&   DA+dM  &2451047370931938176 &52,870 (1219)       &7.83 (0.08)    &0.61 (0.04)    &      H        &   94&6.24      &1                   \\
\\
0133$-$3846&   DA     &5008992762714188800 &16,740 (286)        &8.12 (0.05)    &0.69 (0.03)    &      H        &  117&8.25      &1                   \\
0134$-$4042&   DA+dM  &4960508526175740928 &34,930 (560)        &7.79 (0.07)    &0.56 (0.03)    &      H        &  264&6.74      &1                   \\
0136$-$2010&   DA     &5139880551029408768 & 8470 (123)         &8.64 (0.06)    &1.01 (0.03)    &      H        &   24&9.49      &1                   \\
0137$-$2908&   DA     &5023333658515040128 &24,260 (532)        &7.60 (0.07)    &0.44 (0.03)    &      H        &  158&7.17      &1                   \\
0138$-$2336&   DA     &5039292451317431168 &36,540 (782)        &7.71 (0.10)    &0.52 (0.04)    &      H        &  482&6.65      &1                   \\
\\
0138$-$4014&   DA     &4960499042889705344 &23,540 (530)        &8.05 (0.07)    &0.66 (0.04)    &      H        &  138&7.50      &1                   \\
0140$-$3914&   DA     &4962193390308361728 &23,720 (535)        &8.19 (0.07)    &0.74 (0.04)    &      H        &   58&7.70      &1                   \\
0142$-$3026&   DA     &5022387185162024960 &11,220 (262)        &7.73 (0.12)    &0.46 (0.06)    &      H        &   84&8.51      &1                   \\
0145$-$2211&   DA     &5135466183642594304 &11,390 (263)        &8.17 (0.10)    &0.71 (0.06)    &      H        &   46&8.76      &1, 2                \\
0145$-$7035&   DA     &4687960621811310976 &19,210 (426)        &7.79 (0.07)    &0.51 (0.03)    &      H        &  122&7.68      &3                   \\
\\
\\
0146$-$2500&   DA     &5038054122050564864 &21,810 (873)        &7.51 (0.12)    &0.40 (0.04)    &      H        &  301&7.29      &1                   \\
0149$-$2518&   DB     &5026000176075186560 &16,640 (544)        &7.87 (0.23)    &0.52 (0.12)    &      He       &  147&8.04      &1                   \\
0154$-$2405&   DA     &5122240020833180416 &76,420 (3538)       &7.46 (0.16)    &0.55 (0.05)    &      H        &  545&5.27      &1                   \\
0158$-$1600&   DB     &5147930591051748480 &27,020 (1788)       &8.03 (0.08)    &0.63 (0.04)    &      He       &   68&7.19      &                    \\
0158$-$2242&   DA     &5134814168951616000 &72,940 (3210)       &7.43 (0.15)    &0.53 (0.05)    &      H        &  594&---       &                    \\
\\
0200$-$1243&   DC     &5149836834977282048 & 9350 (39)          &8.05 (0.01)    &0.61 (0.00)    &log H/He=$-$4.4&   24&8.92      &4                   \\
0203$-$1807&   DC     &5138313850039513472 &10,030 (89)         &7.80 (0.02)    &0.47 (0.01)    &log H/He=$-$4.6&   52&8.70      &4                   \\
0205$-$3025&   DA     &5020146620982523008 &18,330 (576)        &8.01 (0.10)    &0.62 (0.06)    &      H        &   87&8.00      &                    \\
0205$-$3635&   DA     &4968128554074903808 &67,060 (1887)       &7.65 (0.10)    &0.58 (0.03)    &      H        &  382&5.87      &1                   \\
0208$-$1520&   DA+dM  &5148232991109667328 &20,890 (390)        &8.14 (0.06)    &0.71 (0.03)    &      H        &   96&7.90      &1                   \\
\\
0208$-$2621&   DA     &5118133276183878400 &33,720 (544)        &8.03 (0.07)    &0.67 (0.04)    &      H        &  250&6.79      &1                   \\
0210$-$3929&   DC     &4964236282912470400 & 8170 (63)          &7.92 (0.01)    &0.53 (0.01)    &log H/He=$-$3.8&   54&8.99      &4                   \\
0212$-$2308&   DA     &5123291871208145792 &27,270 (468)        &8.18 (0.06)    &0.74 (0.04)    &      H        &  184&7.31      &                    \\
0219$-$4049&   DA     &4951125156507440384 &15,760 (295)        &8.09 (0.05)    &0.67 (0.03)    &      H        &  100&8.31      &                    \\
0221$-$2642&   DA     &5118624311204821504 &33,060 (567)        &7.78 (0.08)    &0.55 (0.04)    &      H        &  178&6.80      &                    \\
\\
0222$-$2630&   DA     &5118634309888681600 &24,320 (448)        &8.00 (0.06)    &0.64 (0.03)    &      H        &  108&7.37      &                    \\
0226$-$3255&   DA     &5063539946887524864 &22,710 (551)        &8.15 (0.08)    &0.72 (0.05)    &      H        &   44&7.74      &1                   \\
0235$-$1234&   DA     &5170668766392712448 &32,600 (520)        &8.48 (0.07)    &0.94 (0.04)    &      H        &   70&7.39      &1                   \\
0252$-$3501&   DA     &5049628376014475648 &17,710 (283)        &7.57 (0.05)    &0.41 (0.02)    &      H        &  130&7.69      &1                   \\
0257$-$3339&   DA     &5051294582807658240 &44,570 (1616)       &7.82 (0.15)    &0.59 (0.07)    &      H        &  272&6.44      &                    \\
\\
0300$-$2313&   DA     &5078074837769743744 &24,300 (457)        &8.46 (0.06)    &0.91 (0.04)    &      H        &   79&8.00      &                    \\
0308$-$2305&   DA     &5075443981321647744 &25,940 (450)        &8.62 (0.06)    &1.01 (0.04)    &      H        &   55&8.06      &                    \\
0309$-$2105&   DC     &5102795329494950784 & 9970 (94)          &7.95 (0.02)    &0.55 (0.01)    &log H/He=$-$4.6&   46&8.79      &4                   \\
0309$-$2730&   DA+dM  &5061294228745514752 &61,970 (2533)       &7.21 (0.14)    &0.43 (0.04)    &      H        &  269&---       &1                   \\
0315$-$3313&   DA     &5054287144220982144 &46,130 (1314)       &7.57 (0.11)    &0.50 (0.04)    &      H        &  486&6.17      &                    \\
\\
0320$-$5355&   DA     &4734298439852277376 &32,660 (571)        &8.01 (0.09)    &0.66 (0.05)    &      H        &  118&6.83      &                    \\
0333$-$3500&   DA     &4861136658123791488 &34,180 (518)        &7.95 (0.06)    &0.63 (0.03)    &      H        &  237&6.77      &1                   \\
0335$-$4205&   DA     &4849023441599834112 &44,880 (1697)       &7.74 (0.16)    &0.56 (0.06)    &      H        &  295&6.42      &                    \\
0341$-$6719&   DA     &4667932055438729728 &15,840 (503)        &8.27 (0.08)    &0.78 (0.05)    &      H        &   63&8.43      &                    \\
0400$-$4045&   DA     &4842472688760384384 &10,330 (160)        &7.95 (0.07)    &0.57 (0.04)    &      H        &   45&8.73      &                    \\
\\
0413$-$4029&   DA     &4841120770494891520 &53,110 (2129)       &7.75 (0.15)    &0.58 (0.06)    &      H        &  287&6.21      &                    \\
0420$-$7310&   DA     &4653404070862114176 &18,760 (428)        &8.14 (0.07)    &0.70 (0.04)    &      H        &   84&8.09      &                    \\
0442$-$3523&   DBA    &4867104257483687040 &13,400 (400)        &7.95 (0.19)    &0.56 (0.11)    &log H/He=$-$4.8&   59&8.43      &                    \\
0455$-$2812&   DA     &4880286371109059712 &59,240 (1454)       &7.81 (0.09)    &0.62 (0.03)    &      H        &  125&6.09      &1                   \\
0458$-$3020&   DA     &4876440794831381760 &84,500 (4212)       &7.46 (0.16)    &0.57 (0.05)    &      H        &  681&5.03      &1                   \\
\\
0501$-$2858&   DOZ    &4876967941937370368 &73,750 (3527)       &8.06 (0.20)    &0.71 (0.10)    &      He       &  138&5.89      &1                   \\
1950$-$4314&   DA     &6685284241684914688 &41,310 (1160)       &7.84 (0.12)    &0.59 (0.05)    &      H        &  142&6.53      &1                   \\
1958$-$5008&   DA     &6667323341287601024 &52,070 (1776)       &7.95 (0.13)    &0.67 (0.06)    &      H        &  149&6.25      &                    \\
2000$-$5611&   DA     &6448025712768292992 &42,940 (1278)       &7.74 (0.12)    &0.55 (0.05)    &      H        &  182&6.47      &1                   \\
2001$-$5349&   DA     &6473219235012100096 &17,340 (302)        &7.99 (0.05)    &0.61 (0.03)    &      H        &  108&8.08      &                    \\
\\
\\
2002$-$5405&   DA     &6473244283259625088 &18,180 (298)        &8.02 (0.05)    &0.63 (0.03)    &      H        &  101&8.02      &                    \\
2004$-$3021&   DA     &6749528431218859392 &17,040 (609)        &8.17 (0.11)    &0.72 (0.07)    &      H        &   86&8.26      &                    \\
2017$-$3726&   DA     &6693472717093585664 &19,940 (650)        &7.78 (0.10)    &0.51 (0.05)    &      H        &  178&7.60      &                    \\
2019$-$3438&   DA     &6696596356613398144 &27,010 (599)        &8.29 (0.09)    &0.81 (0.05)    &      H        &  146&7.54      &                    \\
2020$-$4234&   DA     &6679362959252072832 &29,620 (515)        &8.21 (0.07)    &0.76 (0.04)    &      H        &   98&7.13      &1                   \\
\\
2031$-$2745&   DA     &6799465466213872256 &55,960 (1080)       &7.78 (0.07)    &0.60 (0.03)    &      H        &  183&6.16      &                    \\
2033$-$2525&   DBA    &6800380775284148864 &15,230 (356)        &7.94 (0.07)    &0.56 (0.04)    &log H/He=$-$4.8&   61&8.25      &                    \\
2035$-$3337&   DA     &6791907526363085952 &19,650 (411)        &7.75 (0.07)    &0.50 (0.03)    &      H        &   53&7.61      &1                   \\
2035$-$4236&   DA     &6676510104534576512 &69,230 (6699)       &6.52 (0.27)    &0.31 (0.06)    &      H        &  700&---       &3                   \\
2040$-$3914&   DA     &6681773947733560192 &10,560 (165)        &7.85 (0.07)    &0.52 (0.04)    &      H        &   23&8.66      &1, 2                \\
\\
2046$-$3016&   DBA    &6794613321399870080 &11,280 (314)        &7.75 (0.18)    &0.45 (0.09)    &log H/He=$-$5.2&   49&8.53      &                    \\
2100$-$3821&   DA     &6774392546650858624 &16,250 (362)        &8.12 (0.07)    &0.69 (0.04)    &      H        &   78&8.29      &                    \\
2100$-$3840&   DA     &6774367017365076608 &23,930 (656)        &8.07 (0.09)    &0.67 (0.05)    &      H        &  194&7.49      &                    \\
2101$-$3627&   DA+dM: &6776891083744902784 &44,720 (1237)       &7.49 (0.11)    &0.47 (0.04)    &      H        &  349&5.91      &                    \\
2103$-$3947&   DA     &6774018369099513216 & 7780 (149)         &8.23 (0.17)    &0.74 (0.11)    &      H        &   25&9.23      &                    \\
\\
2107$-$2653&   DA     &6802767849386125440 &88,370 (4225)       &7.16 (0.14)    &0.51 (0.03)    &      H        &  577&---       &                    \\
2108$-$4310&   DA+dM  &6580207798067984000 &23,080 (466)        &8.31 (0.07)    &0.82 (0.04)    &      H        &  132&7.92      &                    \\
2109$-$2934&   DC     &6788656957673130112 & 9260 (40)          &7.98 (0.01)    &0.57 (0.00)    &log H/He=$-$4.4&   32&8.89      &4                   \\
2113$-$3705&   DC     &6776236530729179136 & 9590 (95)          &7.83 (0.02)    &0.48 (0.01)    &log H/He=$-$4.4&   71&8.76      &4                   \\
2114$-$3737&   DA     &6583681533257898752 &23,190 (577)        &7.88 (0.08)    &0.57 (0.04)    &      H        &  119&7.38      &1                   \\
\\
2122$-$4643&   DA     &6575491133702739584 &16,130 (276)        &8.28 (0.05)    &0.78 (0.03)    &      H        &   85&8.42      &                    \\
2124$-$4158&   DB     &6578965762246071808 &18,050 (741)        &8.46 (0.23)    &0.89 (0.15)    &      He       &  114&8.42      &                    \\
2133$-$3637&   DA     &6589423148617705856 &24,880 (571)        &7.83 (0.08)    &0.55 (0.04)    &      H        &  118&7.22      &1                   \\
2137$-$3651&   DQ     &6589369272547881856 &12,220 (120)        &8.78 (0.01)    &1.08 (0.01)    &log C/He=$-$0.8&   39&9.16      &                    \\
2137$-$3756&   DA     &6585878288771383296 &20,500 (686)        &8.01 (0.10)    &0.63 (0.06)    &      H        &  145&7.78      &                    \\
\\
2146$-$4320&   DA     &6565941703417200128 &64,130 (2474)       &7.64 (0.14)    &0.57 (0.04)    &      H        &  343&5.92      &1                   \\
2148$-$2910&   DA     &6617454235493798272 &12,480 (232)        &8.05 (0.06)    &0.63 (0.04)    &      H        &   74&8.58      &1, 2                \\
2148$-$2928&   DOZ    &6617395854003241856 &95,040 (13197)      &8.00 (0.35)    &0.72 (0.14)    &      He       &  696&5.56      &                    \\
2151$-$3043&   DA     &6616313457820826496 &28,770 (558)        &8.32 (0.08)    &0.83 (0.05)    &      H        &   69&7.41      &                    \\
2153$-$4156&   DA     &6572442909513784576 &42,590 (2031)       &7.81 (0.21)    &0.58 (0.09)    &      H        &  209&6.49      &1                   \\
\\
2154$-$4342&   DB     &6570892323240774144 &16,180 (513)        &7.86 (0.23)    &0.51 (0.12)    &      He       &   50&8.08      &                    \\
2159$-$4129&   DA     &6571754443436630272 &60,990 (3219)       &7.54 (0.19)    &0.53 (0.06)    &      H        &  270&5.74      &1                   \\
2214$-$3740&   DA     &6598092346829714560 &31,360 (560)        &8.06 (0.09)    &0.68 (0.05)    &      H        &  171&6.90      &1                   \\
2227$-$3246&   DO     &6601723758858306304 &49,700 (978)        &8.01 (0.38)    &0.65 (0.19)    &      He       &  305&6.36      &                    \\
2231$-$2935&   DA     &6608413462479644928 &17,280 (405)        &8.20 (0.07)    &0.73 (0.04)    &      H        &---  &8.26      &5                   \\
\\
2233$-$3529&   DA     &6597106153619823488 &29,350 (535)        &8.06 (0.08)    &0.68 (0.04)    &      H        &  131&7.02      &                    \\
2241$-$4031&   DC     &6545617505854472192 &10,080 (141)        &7.88 (0.03)    &0.52 (0.01)    &log H/He=$-$4.6&   69&8.74      &4                   \\
2251$-$6326&   DA     &6394241555306538240 &18,940 (410)        &7.92 (0.07)    &0.57 (0.04)    &      H        &   46&7.84      &1                   \\
2252$-$4042&   DA+dM: &6546151898569986688 &66,130 (2757)       &7.85 (0.15)    &0.65 (0.06)    &      H        &  911&5.95      &                    \\
2259$-$2646&   DA     &2382502441666663168 &19,620 (529)        &7.82 (0.08)    &0.53 (0.04)    &      H        &   74&7.67      &1                   \\
\\
\\
2259$-$3212&   DA+dM  &6604630592725421440 &85,710 (6686)       &6.99 (0.21)    &0.47 (0.05)    &      H        &  317&---       &                    \\
2306$-$2726&   DA     &2379359556397873664 &19,140 (715)        &7.53 (0.12)    &0.40 (0.04)    &      H        &  163&7.53      &1                   \\
2311$-$2605&   DA     &2379935013296113664 &45,480 (1933)       &8.14 (0.18)    &0.75 (0.10)    &      H        &  315&6.38      &                    \\
2313$-$3303&   DA+dM  &6555925496084361344 &41,560 (971)        &7.41 (0.10)    &0.44 (0.03)    &      H        &  463&5.76      &                    \\
2318$-$2236&   DA     &2385336982642954496 &31,110 (637)        &7.89 (0.11)    &0.59 (0.05)    &      H        &  185&6.90      &                    \\
\\
2322$-$1808&   DA     &2393546245693497728 &22,130 (640)        &7.76 (0.09)    &0.51 (0.04)    &      H        &   93&7.39      &                    \\
2326$-$2226&   DA     &2388347308040329216 &21,100 (1033)       &7.91 (0.15)    &0.58 (0.08)    &      H        &  109&7.60      &                    \\
2329$-$3317&   DA     &2325117757286653440 &20,600 (668)        &7.97 (0.10)    &0.61 (0.06)    &      H        &  148&7.72      &                    \\
2330$-$2113&   DA     &2388953031573382784 &26,700 (517)        &7.63 (0.07)    &0.47 (0.03)    &      H        &  263&7.04      &                    \\
2331$-$4731&   DA     &6528109879126984960 &54,350 (1145)       &7.74 (0.07)    &0.58 (0.03)    &      H        &  106&6.17      &1                   \\
\\
2333$-$1634&   DA     &2395444208921491456 &14,240 (607)        &8.09 (0.07)    &0.66 (0.04)    &      H        &   24&8.44      &                    \\
2334$-$4127&   DB     &6537005065634429312 &21,190 (994)        &8.28 (0.09)    &0.77 (0.06)    &      He       &   75&8.03      &                    \\
2336$-$1842&   DA     &2393875961742886656 & 7710 (126)         &7.39 (0.15)    &0.31 (0.05)    &      H        &   37&8.79      &3                   \\
2336$-$1955&   DA     &2390640129086674560 &20,060 (328)        &8.05 (0.05)    &0.65 (0.03)    &      H        &  145&7.87      &1                   \\
2343$-$1740&   DA     &2394370123500185344 &22,460 (381)        &8.12 (0.05)    &0.70 (0.03)    &      H        &  165&7.72      &1                   \\
\\
2345$-$3940&   DA     &6534665545408513792 &20,650 (328)        &7.95 (0.05)    &0.60 (0.03)    &      H        &  113&7.69      &1                   \\
2347$-$1916&   DA     &2390888829168611968 &27,600 (599)        &7.86 (0.09)    &0.57 (0.04)    &      H        &  150&7.07      &1                   \\
2349$-$2819&   DA     &2334079090586541440 &18,670 (478)        &7.96 (0.08)    &0.60 (0.04)    &      H        &   86&7.91      &1                   \\
2349$-$3627&   DA     &2311200968031379584 &51,490 (1124)       &7.64 (0.08)    &0.54 (0.03)    &      H        &  351&6.18      &1                   \\
2350$-$2448&   DA     &2338275136195124864 &30,560 (608)        &8.41 (0.10)    &0.89 (0.06)    &      H        &   87&7.43      &1                   \\
\\
2352$-$1249&   DA     &2421871039614828672 &44,930 (1519)       &8.00 (0.14)    &0.68 (0.07)    &      H        &  444&6.42      &1                   \\
2354$-$1510&   DA     &2419140746085234688 &37,220 (605)        &7.17 (0.06)    &0.34 (0.02)    &      H        &  347&---       &1, 3                \\
2354$-$3033&   DB     &2326837707005769216 &24,950 (2645)       &7.98 (0.08)    &0.60 (0.04)    &      He       &  156&7.32      &1                   \\
2359$-$3228&   DA     &2313582750735435776 &22,780 (372)        &7.99 (0.05)    &0.62 (0.03)    &      H        &  192&7.51      &1                   \\
\enddata
\tablecomments{
(1) Spectroscopically classified as a white dwarf in the MWDD; 
(2) ZZ Ceti variable; 
(3) Double degenerate binary candidate; 
(4) Physical parameters determined from the {\it Gaia} photometry and parallax; 
(5) No {\it Gaia} parallax available. }
\end{deluxetable}

\clearpage
\bibliography{ms}{}
\bibliographystyle{apj}

\end{document}